\documentclass[aps,floatfix,onecolumn,a4paper,noshowpacs, nofootinbib,superscriptaddress,11pt]{revtex4}
\usepackage{graphicx,float,tikz,makecell}
\usepackage[all]{xy}
\usepackage{physics,bm,amsmath,upgreek,bigints}
\usepackage{amssymb}
\usepackage{color}
\usepackage{epsfig,bm}		
\usepackage{graphicx,epstopdf}
\usepackage{subfigure}
\usepackage{pdfpages}
\usepackage{multirow}
\usepackage{array}
\newcommand{\clt}{\textcolor{black}}
\usepackage{bookmark,textgreek}
\usepackage{hyperref,color,xcolor}
\hypersetup{hidelinks,colorlinks=true,breaklinks=true,urlcolor= blue}
\hypersetup{%
    colorlinks = true,
    linkcolor  = blue,
    citecolor = cyan,
  }

\usepackage{natbib}
\setcitestyle{square,numbers}

\newcommand\orcidroldao{{\href{https://orcid.org/0000-0003-3978-532X}{\orcidicon}}}
\newcommand\orcidwayne{{\href{https://orcid.org/0000-0002-7701-0421}{\orcidicon}}}
\newcommand{\orcidicon}{%
	\begin{tikzpicture}
	\draw[lime, fill=lime] (0,0)
		circle [radius=0.16]
		node[white] {{\fontfamily{qag}\selectfont \tiny ID}};
	\draw[white, fill=white] (-0.0625,0.095)
		circle [radius=0.007];
	\end{tikzpicture}	\hspace{-2mm}
}
\newcommand{\bpartial}{\mathop{\partial\kern -4pt\raisebox{.8pt}{$|$}}}

\newcommand{\bes}{\begin{subequations}}
\newcommand{\ees}{\end{subequations}}
\def\beq{\begin{eqnarray}}

\def\eeq{\end{eqnarray}}
\def\be{\begin{equation}}
\def\ee{\end{equation}}

\begin{document}

\title{Spectroscopy of charmonium-like mesons, heavy-light mesons with charm, AdS/QCD, and configurational entropy}

\author{A. E. Bernardini}
\email{alexeb@ufscar.br}
\altaffiliation[On leave of absence from]{~Departamento de F\'{\i}sica, Universidade Federal de S\~ao Carlos, PO Box 676, 13565-905, S\~ao Carlos, SP, Brasil.}
\affiliation{Departamento de F\'isica e Astronomia, Faculdade de Ci\^{e}ncias da
Universidade do Porto, Rua do Campo Alegre 687, 4169-007, Porto, Portugal.}

\author{W. de Paula\orcidwayne\!\!}
\email{wayne@ita.br}
\affiliation{Instituto Tecnol\'ogico de Aeron\'autica,  DCTA, S\~ao Jos\'e dos Campos, 12228-900, Brazil}

\author{R. da Rocha\orcidroldao\!\!}
\email{roldao.rocha@ufabc.edu.br}
\altaffiliation[(corresponding author).]{}
\affiliation{Federal University of ABC, Center of Mathematics, Santo Andr\'e, 09580-210, Brazil}

\begin{abstract}
Heavy-light-flavor meson resonances with charm, in the $D^0$ and $D^*$ families, and charmonium-like states, in the $\eta_c$ and $\chi_{c1}$ families, are explored and discussed in the AdS/QCD model with four quark flavors. 
The differential configurational entropy is computed and analyzed for these four charmed meson families, also combining 4-flavor AdS/QCD to experimental data for the $D^0$, $D^*$, $\eta_c$, and $\chi_{c1}$ meson families. It makes it possible to predict the mass spectrum of unexplored heavier charmed meson resonances and to identify further charmed meson states reported in PDG.

 \end{abstract}
\pacs{89.70.Cf, 11.25.Mj, 14.40.-n }
\maketitle

\section{Introduction}

The differential configurational entropy (DCE) comprises, along with the differential configurational complexity, a configurational information measure underlying any physical system  \cite{Gleiser:2018kbq,Gleiser:2018jpd}. The construction of the DCE 
is the limit of the canonical Shannon's information entropy for  continuous medium models as long as the information dimension equals zero in the continuous limit. The DCE carries the correlation function  describing the wave modes that constitute the continuous physical system and is a measure of the order in the physical system.  
Ref. \cite{Bernardini:2016hvx} initiated the investigation of AdS/QCD from the point of view of the DCE. Since then, the DCE has been approaching and addressing significant physical aspects regarding hadronic matter in QCD. The DCE makes the holographic AdS/QCD correspondence to merge with experimental data reported in the Particle Data Group (PDG) \cite{pdg}, exploring several features of confined hadrons and corroborating with phenomenological data to good accuracy. The decay width, the stability, and the mass spectrum of hadrons and their heavier resonances have been surveyed by DCE techniques  \cite{Bernardini:2018uuy,Karapetyan:2018oye,Karapetyan:2018yhm,MartinContreras:2020cyg,Braga:2020myi,daRocha:2021imz,Braga:2023qee,Braga:2022yfe,daRocha:2024sjn,Karapetyan:2023sfo}. The DCE-based setup has been fixing key issues and solving relevant questions about hadronic matter.
Besides, the DCE was used to estimate the mass spectrum of higher-excited heavier resonances in some meson families, including  charmonia and bottomonia, hadrocharmonia, hybrid mesons,  quarkonium-like exotica,  tensorial mesons, glueballs,  
 hadronic molecules, odderons, and tetraquarks, among other meson states, including  zero temperature and finite temperature as well \cite{Karapetyan:2021ufz,Karapetyan:2019fst,Karapetyan:2021crv,Ma:2018wtw,MartinContreras:2023oqs,Braga:2020opg,Ferreira:2019inu,daRocha:2021ntm,Braga:2017fsb,Colangelo:2018mrt,DaRocha:2019fjr,Ovalle:2018ans,Braga:2020hhs}. The DCE has been also employed to probe new physical aspects regarding baryon resonances in Refs. \cite{Ferreira:2020iry,daRocha:2022bnk,MartinContreras:2022lxl,Karapetyan:2023kzs,Guo:2024nrf,Karapetyan:2022rpl}, supporting  experimental data reported in PDG \cite{pdg}.  The DCE has been playing an important role in addressing QFT  \cite{Bazeia:2018uyg,Bazeia:2021stz,Feitosa:2024jtr,Barreto:2022ohl,Mvondo-She:2023xel,Cruz:2019kwh,Koliogiannis:2024szd,Lee:2019tod,Lee:2018zmp}.

AdS/QCD is a very convenient paradigm to scrutinize meson resonances in nonperturbative QCD,  where meson states are dual to  gauge fields with flavor in a Yang--Mills theory   \cite{Callebaut:2011ab,Colangelo:2008us,Csaki,Branz:2010ub}. Among the bottom-up approaches of AdS/QCD, the  soft-wall AdS/QCD can be implemented when a dilaton field couples to Einstein gravity. In this setup, Regge trajectories of mesons are well described and confinement is introduced dynamically, besides fitting QCD phenomenology with good accuracy \cite{Karch:2006pv,Gursoy:2007cb,dePaula:2008fp,Li:2013oda,He:2013qq,Bartz:2014oba,Bartz:2018nzn,Ghoroku:2005vt,Vega:2018dgk,dePaula:2009za,Ballon-Bayona:2017sxa,Afonin:2022hbb,Ballon-Bayona:2023zal}.

Charm meson spectroscopy has recently 
 resurged as a consequence of the production of several charm and charm-strange resonances by the $B$-factory experiments BaBar, LHCb, and Belle \cite{BaBar:2010zpy,LHCb:2024vfz,LHCb:2015eqv,LHCb:2019juy}.  Matching theoretical predictions to experimentally measured data yielded significant efforts to engender a quark constituent model approach to the spectroscopic data for dozens of new meson resonances. Concomitantly, higher energy beams at the LHC
and higher luminosity beams at the SuperKEKB collider suggest that new meson resonances will be produced and detected.  The quark constituent model has been fruitful in addressing and calculating the properties
of excited charm and charm-strange meson resonances.
Charmed mesons have also been investigated in Ref. \cite{Ballon-Bayona:2017bwk}, with relevant experimental signatures in the quark-gluon plasma \cite{Torrieri:2010py,Machado:2013rta}. Charmonium dissociation has been scrutinized in holographic QCD \cite{Dudal:2014jfa,Guo:2024qiq}. There is considerable interest in
exploring the interactions of charmed hadrons with 
light-flavor hadrons, with particular interest 
in heavy-light-flavor charmed mesons. The experimental detection of the charm quark was an important validation of the Standard
Model (SM), as its existence and mass scale were predicted based on 
low-energy kaon experiments. Several features distinguish charmed hadrons from those with other flavors.
While their mass is around 2 GeV order, corresponding to nonperturbative
hadronic physics, theoretical methods developed for heavy quarks can  still be applied. On the other
hand, recent advances in unquenched lattice QCD simulations paved the way for charm data to be used to probe the Yukawa sector of the SM. Charm transitions provide low-energy signatures of new physics at the TeV scale, whose models predict signals for CP-violation much larger than the predicted ones by SM \cite{Artuso:2008vf}. 
Experimental data yield a well-established model to describe $D^0$ and $D^*$ meson resonances, which have
been observed by $B$-factories and LHCb, which raises
much interest amid newly observed charmed resonances.
Several new excited states of heavy-light mesons were discovered, some of which have rather unexpected
properties \cite{Godfrey:2015dva}.
The LHCb collaboration has detected CP violation in the $D^0$ charm meson, corroborating the pattern underlying the CP violation in the SM, adding new ramifications  \cite{LHCb:2022jaa,Ebert:2009ua}. 
For the charmed counterparts $D^*(2007)$,  $D^*_1(2600)$, and $D^*_1(2760)$,  most of the theoretical
approaches, including QCD sum rules,  different quark constituent models, and lattice QCD, imply masses of the $0^+$ and $1^+$  $P$-wave $c\bar{s}$ states significantly heavier by $\sim200$
MeV than the measured ones.
Distinct theoretical approaches for this mismatch between theory and experiments were addressed, 
 including proposing these mesons as chiral partners of $0^-$ and $1^-$ states \cite{Lewis:2000sv}. Nevertheless, a more universal and deep understanding of their
nature is still lacking. 

Quarkonium is a multiscale system that can
probe all regimes of QCD.
At high energies, a perturbative expansion in the strong
coupling constant is feasible, as nonperturbative effects dominate at low energies. Hence, heavy quarkonium comprises an ideal
laboratory for exploring the interplay between nonperturbative and perturbative QCD. Quarkonia have been probed by experiments by the BESIII collaboration, $B$-factories, CLEO-c, HERA, Tevatron,  LHC, RHIC, FAIR, the Super Flavor and Tau-Charm factories, JLab, the ILC, and beyond. Such experiments unleashed newly-found conventional meson resonances and some unexpected ones, like the $\chi_{c1}(3872)$ and other quarkonium-like
states \cite{Brambilla:2010cs}. Decays of the $\chi_{c1}$ $P$-wave charmonium states are considered an ideal laboratory to test QCD \cite{BESII:2011hcd}.

It is a challenge to identify new
charmed meson resonances. In this work, we aim to deeply investigate charmonium-like states and heavy-light-flavor charmed mesons. AdS/QCD with four quark flavors ($u$, $d$, $s$, and $c$) will be combined with the DCE as a configurational information measure. The families $D^0$ and $D^*$, consisting of heavy-light mesons with charm, and  the charmonium-like families $\eta_c$ and $\chi_{c1}$  will be delved into, in this hybrid setup. 
Subsequent to Ref. \cite{Erlich:2005qh}, which proposed  a chiral symmetry-breaking holographic realization of the hard-wall AdS/QCD, Ref. \cite{Ballon-Bayona:2017bwk} reported an extension to AdS/QCD with four flavors,  embracing both flavor  and chiral symmetry-breaking. The protocol used for implementing it used a Kaluza--Klein expansion of the AdS bulk fields to yield an effective four-dimensional action regulating the light-flavor and heavy-light-flavor meson resonances. This approach describes the decay constants with good accuracy, as well as the mass spectrum of charmed mesons. Another approach to describe heavy-light-flavor mesons and charmonium was proposed in Ref. \cite{Chen:2021wzj}, by introducing the scalar field $\mathsf{H}$ to the Einstein-dilaton system.

This apparatus will be employed in this work for computing the  DCE of the $D^0$, $D^*$, $\eta_c$, and $\chi_{c1}$ meson families as a function of the principal quantum number $n$, whose interpolation function is known as the  DCE-Regge trajectories of the first  kind  \cite{Bernardini:2018uuy}.  \clt{The meson mass spectrum of heavy-light-flavor mesons and charmonium, to be used in this work, was already computed in Ref. \cite{Chen:2021wzj}}. Moreover, when the DCE is expressed as a function of the experimental mass spectra of the $D^0$, $D^*$, $\eta_c$, and $\chi_{c1}$ meson families in PDG \cite{pdg}, the interpolation function is called DCE-Regge trajectories of the second  kind. 
The mass spectra of the next generation of these charmed meson resonances can be obtained when the two kinds of Regge trajectories are extrapolated for higher values of $n$.  
Extrapolating the DCE-Regge trajectories yields the mass spectrum of the resonances with $n>4$, in the heavy-light-flavor $D^0$ meson family, when the interpolation formul\ae\; of the DCE-Regge trajectories of the first  and second kinds are worked out for the states $D^0$, $D^0(2550)$, and $D^0(3000)$ reported in the PDG \cite{pdg}. An analogous procedure leads the heavy-light-flavor  $D^*$ meson family for $n>4$ to have their higher excitations to attain a specific mass spectrum, based on the experimental data of $D^*(2007)^0$, $D^*_1(2600)^0$, and $D^*_1(2760)^0$ mesons, respectively corresponding to the meson states in this family with $n=1,2,3$.   
 The properties of $2S$ and $1D$ heavy-light-flavor charmed meson resonances have been
investigated for a long time. Nonetheless, such a higher excited
heavy-light-flavor state has not been evidenced yet, due to the lack of experimental
data \cite{pdg}. Some heavier, higher-excited, charmed meson resonances were reported, though most of
them have not yet been pinned down. The DCE approach of 4-flavor AdS/QCD is a useful one to systemically scrutinize the possible $2S$ and $1D$ charmed
and charmed-strange meson resonances. 
Regarding the $\eta_c$  charmonium-like meson family, values $n>3$ can be studied 
from the $\eta_c(1S)$ and $\eta_c(2S)$ states, revealing the mass spectrum of the $\eta_c(3S)$ and $\eta_c(4S)$ charmonium-like states. Finally, one can determine the mass spectrum of the next generation of heavier resonances in the $\chi_{c1}$ 
 charmonium family, from the mass spectrum of the detected already states $\chi_{c1}(1P)$,  $\chi_{c1}(3872)$, $\chi_{c1}(4140)$, $\chi_{c1}(4274)$, and $\chi_{c1}(4685)$ reported in PDG \cite{pdg}, by using the DCE.

This paper is organized as follows: Sec. \ref{sec1}   introduces the main ingredients of the bottom-up 4-flavor AdS/QCD.  The mass spectra of  heavy-light-flavor meson families $D^0$ and $D^*$, and the $\eta_c$ and $\chi_{c1}$ charmonium-like meson families, are obtained as eigenvalues occurring in systems of Schr\"odinger-like equations,  which are read off the equations of motion obtained from the Euler--Lagrange equations applied to the action functional regulating the $D^0$, $D^*$, $\eta_c$, and $\chi_{c1}$  meson families, by suitably fixing some physical parameters.  
 Sec. \ref{sec2} approaches the DCE paradigm. The DCE for the heavy-light-flavor charmed mesons $D^0$ and $D^*$, and for the charmonium-like families $\eta_c$ and $\chi{c1}$,  is then computed  firstly as a function of the principal quantum number $n$, yielding DCE-Regge trajectories of the first kind. After the DCE of the $D^0$, $D^*$, $\eta_c$, and $\chi_{c1}$  charmed meson families is analyzed as a function of the respective mass spectra of resonances in these charmed meson families, whose interpolation originates DCE-Regge trajectories of the second kind. The mass spectra of the next generation of mesonic resonances in the $D^0$, $D^*$, $\eta_c$, and $\chi_{c1}$ charmed meson families are then acquired for extrapolation of the DCE-Regge trajectories when values of the principal quantum number $n$ which are higher than the ones indexing the resonances in the $D^0$, $D^*$, $\eta_c$, and $\chi_{c1}$ already detected and reported in PDG.  
Conclusions and closing remarks are addressed in Sec. \ref{iv}.

\section{4-flavor AdS/QCD}
\label{sec1}
Refs. \cite{Ballon-Bayona:2017bwk,Chen:2021wzj,Momeni:2020bmy} established  an  AdS/QCD model for four quark flavors. In this setup, the ground states as well as higher-excitation meson states composed of light-flavor and heavy-light-flavor quarks can be consistently constructed. 
This holographic framework will be used here to study heavy-light-flavor mesons and charmonium-like states.
For a hard-wall cutoff, $z_{\textsc{m}}$, the background of AdS/QCD model with four quark flavors is driven by the 5-dimensional action functional \cite{Chen:2021wzj}
	\begin{eqnarray}\label{action}
		\!\!\!\!\!\!\!\!\!\!\!\!S \!&\!=\!&\! -\int_\upepsilon^{z_{\textsc{m}}}\! d^5 x \sqrt{-g}  e^{-\phi}\, {\rm tr}\,\, \left[ |D\mathsf{X}|^{2}  \!-\! 3 |\mathsf{X}|^2 \!+\!|D\mathsf{H}|^2 \!-\! 3 |\mathsf{H}|^2+\frac{1}{4 g_5^2} \left(R^{MN} \mathsf{R}_{MN} + \mathsf{L}^{MN} \mathsf{L}_{MN} \right)
		\right].
	\end{eqnarray}
The 5-dimensional coupling constant $g_5$ can be associated with the number of
colors $N_c$ in the dual theory on the AdS boundary by the expression  $g_5^2=12\pi^2/N_c$, as the gauge group in QCD is SU($N_c$)  \cite{Erlich:2005qh}. 
The covariant derivatives of the fields $\mathsf{X}$ and $\mathsf{H}$ read  
\begin{subequations}
\beq
D_{M} \mathsf{X}&=&\partial_M \mathsf{X} -i\mathsf{L}_M \mathsf{X} +i \mathsf{X} \mathsf{R}_M,\\
D_M \mathsf{H} &=& \partial_M \mathsf{H} - i {V_M^{15}} \mathsf{H} - i \mathsf{H} {V_M^{15}},
\eeq
\end{subequations} 	whereas the Yang--Mills field strengths read 
\begin{subequations}
\beq
\mathsf{L}_{MN} &=& \partial_{[M} \mathsf{L}_{N]} - i \left [\mathsf{L}_M , \mathsf{L}_N \right],\\ 
\mathsf{R}_{MN} &=& \partial_{[M} \mathsf{R}_{N]}- i \left [ \mathsf{R}_M , \mathsf{R}_N \right ], 
\eeq
\end{subequations}
 for   
\beq\mathsf{L}_M=\mathsf{L}_M^a\mathsf{t}^a,\qquad \qquad\mathsf{R}_M=\mathsf{R}_M^a\mathsf{t}^a, 
\eeq {being $\mathsf{t}^a$ ($a=1,\ldots, N_f^2-1 = 15$) the generators of the SU(4) Lie group, associated with $N_f=4$ quark flavors. The  SU(4) generators are
normalized by the trace condition tr $(\mathsf{t}^a \mathsf{t}^b) = \frac12\delta^{ab}$ and are related to the Gell--Mann matrices $\lambda^a$ by
$\mathsf{t}^a = \frac12\lambda^a$}. 
The fields in Eq. (\ref{lra}) respectively transform as the left and right chiral fields
under the SU(4)$_L \;\times$  SU(4)$_R$ chiral symmetry group of 4-flavor AdS/QCD. 
One can  express \beq\label{lra}
\mathsf{R}_M = \frac{\sqrt{2}}{2}\left(V_M-A_M\right),\qquad\qquad \mathsf{L}_M = \frac{\sqrt{2}}{2}\left(V_M + A_M\right),\eeq 
taking into account the vector and the axial-vector fields.  {The scalar field $\mathsf{H}$ is outlined in the action functional (\ref{action}) to account for the gap between the masses of light-flavor and heavy-flavor quarks, also emulating the precise separation between the light-flavor and heavy-flavor branes, in the $D_4$-$D_8$ Sakai--Sugimoto formalism \cite{Liu:2016iqo}. 

In 4-flavor AdS/QCD, the scalar field $\mathsf{X}$ breaks the SU(4)$_L\; \times$  SU(4)$_R$ 
 symmetry of 4-flavor QCD to SU(4)$_V$, with residual symmetry yielding the equations of motion governing the light-flavor vector meson resonances to have the same form as the ones regulating the charmonium and charmonium-like meson families \cite{Chen:2021wzj}. The SU(4) flavor symmetry breaking was explored in $D$-meson couplings to light-flavor hadronic states in Ref. \cite{Fontoura:2017ujf}.  
The $\mathsf{H}$ scalar field is also responsible for breaking the residual SU$(4)_V$ gauge symmetry to the SU$(3)_V$ Lie group.}  
The scalar field $\mathsf{X}$ implements, in addition, a well-defined mapping to the $\langle\bar{q}_\mathsf{R}q_\mathsf{L}\rangle$ operator, on the AdS boundary. {In contrast, the gauge fields components $\mathsf{L}_M^a$ and $\mathsf{R}_M^a$ correspond to the $\langle\bar{q}_{\mathsf{L},\mathsf{R}}\gamma_\mu \mathsf{t}^aq_{\mathsf{L},\mathsf{R}}\rangle$ operator, where $\gamma^\mu$ denote the Dirac gamma matrices}  \cite{Erlich:2005qh}. 
{Chiral symmetry SU(4)$_L \;\times \;$SU(4)$_R$ is described in terms of the left and right current densities
\beq
j_{\mathsf{L},\mathsf{R}}^{\mu , a} = \bar q_{\mathsf{L},\mathsf{R}} \gamma^\mu \mathsf{t}^a q_{\mathsf{L},\mathsf{R}}. 
\eeq
Chiral symmetry can be broken by the presence
of the operator $\bar q q = \bar q_\mathsf{R} q_\mathsf{L} + \bar q_\mathsf{L} q_\mathsf{R}$ and the vector and axial currents are given by $j^{\mu,a}_{V/A}= j^{\mu, a}_\mathsf{R} \pm j^{\mu, a}_\mathsf{L}$. The symmetry
emerging from the conservation of the current density $j^{\mu, a}_V$ in the vector sector is known as isospin symmetry \cite{Ballon-Bayona:2017bwk}.}

Besides, the non-vanishing VEV of $\mathsf{X}$ promotes the chiral symmetry breaking in light-flavor quarks. 
The 4-flavor AdS/QCD model merges both the soft-wall, through the standard quadratic dilaton \cite{Karch:2006pv}, \beq\label{dila}
\phi(z)=\mu^2z^2, 
\eeq  and the hard-wall implemented by the cutoff $0<z<z_{\textsc{m}}$ \cite{Polchinski:2001tt}. It yields the simultaneous description of the mass spectra of meson resonances with light-flavor, heavy-flavor, and heavy-light-flavor constituent quarks.  The scalar field $\mathsf{X}$ expressed by
	\begin{eqnarray}\label{eqkk}
		\mathsf{X} = e^{i\uppi^b\mathsf{t}^b} \, \mathsf{X}_0 \, e^{i\uppi^a\mathsf{t}^a},
	\end{eqnarray}
where 
\beq
\mathsf{X}_0={\rm diag}\left[{\scalebox{.87}{$\upsilon$}}_\ell(z),{\scalebox{.87}{$\upsilon$}}_\ell(z),{\scalebox{.87}{$\upsilon$}}_s(z),{\scalebox{.87}{$\upsilon$}}_c(z)\right],\label{x0}\eeq 
where the indices in Eq. (\ref{x0}) denote  light (up/down) ($\ell$), strange ($s$), and charm ($c$) quarks, whose asymptotic behavior resembles a field with conformal scaling dimension $\Delta=3$ with UV boundary limit  given by 
\clt{\beq\label{eqk}
{\scalebox{.87}{$\upsilon$}}_{\ell,s,c}(z)\approx M_{\ell,s,c}z+\Upsigma_{\ell,s,c}z^3,
\eeq}  where $\Upsigma^{\rho\sigma} = \langle \bar{q}^\rho q^\sigma\rangle$ \cite{Karch:2006pv}.  
{The 5-dimensional (squared) mass of the field $\mathsf{X}$ is fixed to the expression  $m^2 = -3$, to be consistent with the conformal scaling dimension $\Delta=3$ of the dual operator $\bar q_\mathsf{R} q_\mathsf{L}$.\! }\!\!\!\!\!
Choosing the chiral condensate $\Upsigma= \textsc{diag}(\Upsigma_\ell, \Upsigma_\ell, \Upsigma_s, \Upsigma_c)$ to be diagonal
means that the diagonal subgroup SU(3)$_{\textsc{diag}}$  remains unbroken, as expected, and the leading-order term in $\mathsf{X}$ can be naturally interpreted as the quark mass matrix
{$M^{ab}=\textsc{diag}(m_\ell, m_\ell, m_s, m_c)$. It is worth emphasizing that SU(2) isospin 
symmetry is assumed in the light-quark sector, meaning that 
$m_\ell = m_d \approxeq m_u$ and
$\Upsigma_\ell=\Upsigma_d \approxeq \Upsigma_u$, as in 
Refs. \cite{Ballon-Bayona:2017bwk, Chen:2021wzj}, 
which represents an accurate, approximate symmetry in QCD. There is no formal restriction, a priori, on the formation of a charm quark condensate. However, its contribution is much smaller than that of light  quarks consisting of the up, down, and strange quarks, since the masses of these light-flavor quarks are much smaller than the QCD scale ($\{m_u, m_d, m_s\}\ll\Lambda_{\textsc{QCD}}$). The chiral symmetry breaking is much more relevant in light quarks, and non-perturbative effects significantly increase their mass. For heavy quarks, including the charm, $m_c > \Lambda_{\textsc{QCD}}$, and non-perturbative effects are no longer as relevant as in the case of the condensate of light quarks. Also, as in the energy scale where heavy quarks are more relevant, one must take into account the asymptotic freedom, and the dynamical effects of non-perturbative mass generation present in the confinement scale are much less relevant, as most of the effective mass of heavy quarks comes from the Higgs mechanism. The gluon condensate is more relevant at this scale to affect the effective mass of heavy quarks. 
Some works consider that charm quark condensates have a non-zero effect. One can see, e.g., Ref. \cite{Narison:2010cg} presenting the heavy quark condensate contribution, absorbed into the gluon condensate. To sum up, although the existence of the charm quark condensate is not forbidden, it does contribute just with a minor experimental effect and can be even negligible, when compared to the light quark condensates. The existence of the charm condensate is not prohibited, but its effect is highly suppressed.}
 
It acts as the source of a deformation, 
implemented by the quark mass term, 
of the action of the CFT on the AdS boundary. 
Furthermore, as the
$\mathsf{H}$ scalar encodes the eﬀect of the charm quark mass, one can express 
 \beq
\mathsf{H}={\rm diag}\left[0,0,0,h_c(z)\right],\label{hc}\eeq \clt{where $h_c(z)\approx m_c z$ at the UV boundary} \cite{Chen:2021wzj}. 
The Poincar\'e half-space patch metric equips the  AdS bulk and is given by 
	\begin{eqnarray}\label{eq.1}
		ds^2=\clt{g_{AB}dx^Adx^B} = \frac{L^2}{z^2}\left(dz^2+\upeta_{\mu\nu}dx^{\mu}\,dx^{\nu}\right),
	\end{eqnarray}
with $A,B = 0,1,2,3,5$ denoting indexes labeling the AdS bulk coordinates, $\mu,\nu=0,1,2,3$,  for $\upeta_{\mu\nu}=\rm{diag}[-1,1,1,1]$.
\clt{The boundary of AdS is at $z=0$.} Hereupon, the value $L=1$ will be regarded, for conciseness. 	To derive the masses of meson resonances \clt{and the 3- and 4-point coupling constants}, the action Eq. (\ref{action}) can be expanded to \clt{order four}\footnote{\clt{Here the order four  refers to fluctuations.}}, as 
	\begin{eqnarray}
		S=S^{(0)}+S^{(2)}+S^{(3)}+S^{(4)},\label{act1}
	\end{eqnarray}
	where
	\begin{eqnarray}
		S^{(0)}&=&-\int_0^{z_{\textsc{m}}} d^5x\;\frac{e^{-\phi(z)}}{z^3}~\Bigg\{2{\scalebox{.87}{$\upsilon$}}_\ell^\prime(z) {\scalebox{.87}{$\upsilon$}}_\ell^\prime(z)+{\scalebox{.87}{$\upsilon$}}_s^\prime(z){\scalebox{.87}{$\upsilon$}}_s^\prime(z)+{\scalebox{.87}{$\upsilon$}}_c^\prime(z){\scalebox{.87}{$\upsilon$}}_c^\prime(z)+h_c^{\prime2}(z)\nonumber\\
		&&\qquad\qquad\qquad\qquad\qquad\qquad+\frac{m_5^2}{z^2}\bigg(2{\scalebox{.87}{$\upsilon$}}^2_\ell(z)+{\scalebox{.87}{$\upsilon$}}^2_s(z)+{\scalebox{.87}{$\upsilon$}}^2_c(z)+h_c^2(z)\bigg)\Bigg\},\label{act2}\\
		S^{(2)}&=&-\int_0^{z_{\textsc{m}}} d^5x~\frac{e^{-\phi(z)}}{z}\Bigg\{\upeta^{MN}\frac{1}{z^2}\bigg(\zeta_M^a\zeta_N^b\,\mathsf{M}_A^{ab}-V_M^aV_N^b\mathsf{M}_V^{ab}+V_M^{15}V_N^{15}\mathsf{m}_V^{15,15}\bigg)\nonumber\\
		&&\qquad\qquad\qquad\qquad\qquad\qquad\qquad\qquad+\frac{1}{4g_5^2}\upeta^{MP}\upeta^{NQ}\bigg(V_{MN}V_{PQ}+A_{MN}A_{PQ}\bigg)\Bigg\},\label{act3}\\
		\!\!S^{(3)}\!&\!=\!&\!-\int_0^{z_{\textsc{m}}} d^5x\frac{e^{-\phi(z)}}{z}\Bigg\{\frac{\upeta^{MN}}{z^2}\bigg(2V_M^a\upzeta_N^{bc}\mathsf{h}^{abc}\!-\!2\zeta_M^aV_N^b\uppi^cg^{abc}\bigg)\nonumber\\&&\qquad\qquad\qquad\qquad\qquad\qquad\qquad\!+\frac{\upeta^{MP}\upeta^{NQ}}{2g_5^2}\left(V_{MN}^aV_p^{(b}V_q^{c)}\!+\!A_{MN}^aV_P^{(b}A_Q^{c)}\right)\mathsf{f}^{bca}\Bigg\},\label{act4}\\
		S^{(4)}&=&-\int_0^{z_{\textsc{m}}} d^5x\frac{e^{-\phi(z)}}{z}\Bigg\{\frac{\upeta^{MN}}{z^2}\bigg[\zeta_M^a\mathring{\upzeta}_N^{bcd}\mathsf{l}^{abcd}+V_M^aV_N^b\uppi^c\uppi^d\left(\mathsf{h}^{abcd}-\mathsf{g}^{acbd}\right)+\upzeta_M^{ab}\upzeta_N^{cd}\mathsf{k}^{abcd}\bigg]\nonumber\\
		&&\qquad\qquad\qquad\qquad+\frac{1}{4g_5^2}\upeta^{MP}\upeta^{NQ}\left(V_M^aV_N^bV_P^cV_Q^d
		+V_{(P}^cV_{(Q}^dA_{M)}^aA_{N)}^b+A_M^aA_N^bA_P^cA_Q^d\right.
		\nonumber\\
		&&\left.\qquad\qquad\qquad\qquad\qquad\qquad\qquad+2V_M^aA_N^bV_P^cA_Q^d+2A_M^aV_N^bV_P^cA_Q^d\right)\mathsf{f}^{abcd}\Bigg\},\label{act5}
	\end{eqnarray}
	where 
	\begin{subequations}\beq
	\zeta_M^a&=&\partial_M\uppi^a-A_M^a,\\
	\upzeta_M^{ab}&=&\frac{1}{2}\partial_M(\uppi^a\uppi^b)-A_M^a\uppi^b,\\
	\mathring{\upzeta}^{bcd}&=&\left[A_N^b\uppi^c\uppi^d-\frac{1}{3}\partial_N\left(\uppi^b\uppi^c\uppi^d\right)\right].
	\eeq
	\end{subequations}
	The $\upeta^{MN}$ tensor denotes the 5-dimensional Minkowski spacetime metric, $V_{MN}=\partial_{[M}V_{N]}$, $A_{MN}=\partial_{[M}A_{N]}$, and 
	\begin{eqnarray}
		&\mathsf{M}_A^{ab}={\rm tr}\,\left(\{\mathsf{t}^a,\mathsf{X}_0\}\{\mathsf{t}^b,\mathsf{X}_0\}\right),\qquad 
		&\mathsf{M}_V^{ab}={\rm tr}\,\left([\mathsf{t}^a,\mathsf{X}_0][\mathsf{t}^b,\mathsf{X}_0]\right),\label{t11}\\
		&\mathsf{m}_V^{15,15}={\rm tr}\,(\{H,\mathsf{t}^{15}\}\{H,\mathsf{t}^{15}\}),\qquad &\mathsf{f}^{abcd}=\mathsf{f}^{eab}\mathsf{f}^{ecd}\label{t12}\\
		& \textcolor{black}{\mathsf{h}^{abc} = i~{\rm tr}\,\left([\mathsf{t}^a,\mathsf{X}_0]\{\mathsf{t}^b,\{\mathsf{t}^c,\mathsf{X}_0\}\} \right) },\qquad 
		&\mathsf{g}^{abc}=i~{\rm tr}\,(\{\mathsf{t}^a,\mathsf{X}_0\}[\mathsf{t}^b,\{\mathsf{t}^c,\mathsf{X}_0\}]),\label{t13}\\
                 &\mathsf{h}^{abcd}={\rm tr}\,\left([\mathsf{t}^a,\mathsf{X}_0][\mathsf{t}^b,\{\mathsf{t}^c,\{\mathsf{t}^d,\mathsf{X}_0\}\} ]\right),\qquad 
                 &\mathsf{g}^{abcd}={\rm tr}\,\left([\mathsf{t}^a,\{\mathsf{t}^b,\mathsf{X}_0\}][\mathsf{t}^c,\{\mathsf{t}^d,\mathsf{X}_0\}]\right),\label{t14}\\
		&\textcolor{black}{\mathsf{l}^{abcd}={\rm tr}\,\left(\{\mathsf{t}^a,\mathsf{X}_0\}\{\mathsf{t}^b,\{\mathsf{t}^c,\{\mathsf{t}^d,\mathsf{X}_0\} \}\}  \right)} ,\qquad 
		&\textcolor{black}{\mathsf{k}^{abcd}={\rm tr}\,\left(\{\mathsf{t}^a,\{\mathsf{t}^b,\mathsf{X}_0\}\} \{\mathsf{t}^c,\{\mathsf{t}^d,\mathsf{X}_0\} \}\right) },\label{t15}
	\end{eqnarray}
	where the structure constants $\mathsf{f}^{abc}$ of the SU(4) Lie group are taken into account.

In the 4-flavor AdS/QCD holographic model, the fields can be written in terms of the meson fields as follows \cite{Ballon-Bayona:2017bwk, Chen:2021wzj}:
	\begin{eqnarray}
		\vspace*{-0.5cm}\!\!\!\!\!\!\!\!\!\!\!\!V &=&  \mathsf{V}^a \mathsf{t}^a  = \frac{1}{\sqrt 2}
		\left ( \begin{matrix}
			\frac{\sqrt{2}}{2}${\scalebox{.985}{$\rho^0$}}$ + \frac{\sqrt{6}}{6}{\scalebox{.985}{$\omega'$}} + \frac{\sqrt{3}}{6}{\scalebox{.985}{$\psi$}}  &  {\scalebox{.985}{$\rho^{{+}}$}}  &  {\scalebox{.985}{$K^{{*+}}$}}  &  {\scalebox{.985}{$\bar D^{{*0}}$}}  \\
			\rho^{{-}}   & -\frac{\sqrt{2}}{2}${\scalebox{.985}{$\rho^0$}}$ 
			+ \frac{\sqrt{6}}{6}{\scalebox{.985}{$\omega'$}}  
			+ \frac{\sqrt{3}}{6}{\scalebox{.985}{$\psi$}}   
			&  {\scalebox{.985}{$K^{{*0}}$}}  &  D^{{*-}}  \\
			{\scalebox{.985}{$K^{{*-}}$}}  &  {\scalebox{.985}{$\bar{K}^{{*0}}$}}  & - \frac{\sqrt{6}}{3} {\scalebox{.985}{$\omega'$}} + \frac{\sqrt{3}}{{6}} {\scalebox{.985}{$\psi$}} &  {\scalebox{.985}{$D_s^{{*-}}$}}  \\
			{\scalebox{.985}{$D^{{*0}} $}}&  {\scalebox{.985}{$D^{{*+}} $}}  &  {\scalebox{.985}{$D_s^{{*+}}  $}} & - \frac{\sqrt{3}}{2} {\scalebox{.985}{$\psi$}}
		\end{matrix} \right ), 	\end{eqnarray}
	\begin{eqnarray}
		\vspace*{-0.5cm}\!\!\!\!\!\!\!\!\!\!\!\!A \!&\!=\!&\!  {\mathsf{A}}^a \mathsf{t}^a  \!=\!  \frac{1}{\sqrt 2}\! \left(\begin{array}{cccc}
			\frac{\sqrt{2}}{2}{\scalebox{.985}{$a^0_1$}}  \!+\! \frac{\sqrt{6}}{6}{{\scalebox{.985}{$\eta_c$}}}\!+\!\frac{\sqrt{3}}{6}{\scalebox{.985}{$\chi_{c1}$}}
			&  {\scalebox{.985}{$a^{{+}}_1$}} & {\scalebox{.985}{$K_{1}^{{+}}$}}& {\scalebox{.985}{${\scalebox{.985}{$\bar{D}_{1}^{{0}}$}} $}}\\
			{\scalebox{.985}{$a_1^{{-}}$}}&-\frac{\sqrt{2}}{2} { {\scalebox{.985}{$a^{{0}}_1$}}} \!+\!\frac{\sqrt{6}}{6}{{\scalebox{.985}{$\eta_c$}}}\!+\!\frac{\sqrt{3}}{6}{\scalebox{.985}{$\chi_{c1}$}}
			& {\scalebox{.985}{$K_{1}^{{0}} $}}& {\scalebox{.985}{$D_{1}^{{-}}$}}\\
			{\scalebox{.985}{$K_{1}^{{-}}$}}&{\scalebox{.985}{$ \bar{K}_{1}^{{0}}$}}  & -\frac{\sqrt{6}}{3}{\scalebox{.985}{$\eta_c$}}\!+\!\frac{\sqrt{3}}{6}{\scalebox{.985}{$\chi_{c1}$}}
& {\scalebox{.985}{$D_{s1}^{{-}}$}}\\
			{\scalebox{.985}{${D}_{1}^{{0}}$}}&{\scalebox{.985}{$D_{1}^{{+}}$}}&{\scalebox{.985}{$D_{s1}^{{+}}$}}&- \frac{\sqrt{3}}{2} {\scalebox{.985}{$\chi_{c1}$}}
		\end{array}
		\right), \\
		\vspace*{-0.5cm}\!\!\!\!\!\!\!\!\!\!\!\!\uppi &=& \uppi^a \mathsf{t}^a = \frac{1}{\sqrt 2}
		\left (\begin{matrix}
				\frac{\sqrt{2}}{2}{\scalebox{.985}{$\pi^0$}}  + \frac{\sqrt{6}}{6}{{\scalebox{.985}{$\eta$}}}+\frac{\sqrt{3}}{6}{\scalebox{.985}{$\eta_{c}$}} & {\scalebox{.985}{$ \pi^{{+}}$}}  &  {\scalebox{.985}{$K^{{+}} $}}  &  {\scalebox{.985}{$\bar{D}^{{0}}$}}  \\
			{\scalebox{.985}{$\pi^{{-}}$}}   &-\frac{\sqrt{2}}{2}{\scalebox{.985}{$\pi^0$}}  + \frac{\sqrt{6}}{6}{{\scalebox{.985}{$\eta$}}}+\frac{\sqrt{3}}{6}{\scalebox{.985}{$\eta_{c}$}}  &  {\scalebox{.985}{$K^{{0}} $}} & {\scalebox{.985}{$ D^{{-}} $}} \\
			{\scalebox{.985}{$K^{{-}}$}}  &  {\scalebox{.985}{$\bar K^{{0}}$}}  &  -\frac{\sqrt{6}}{3}{\scalebox{.985}{$\eta$}}+\frac{\sqrt{3}}{6}{\scalebox{.985}{$\eta_{c}$}}  &  {\scalebox{.985}{$D_s^{{-}} $}} \\
			{\scalebox{.985}{$D^{{0}}$}} &  {\scalebox{.985}{$D^{{+}}$}}  &  {\scalebox{.985}{$D_s^{{+}}$}}  & - \frac{\sqrt{3}}{2} {\scalebox{.985}{$\eta$}}_c
		\end{matrix} \right ).\label{m111}
	\end{eqnarray}
	Ref. \cite{Eshraim:2014eka} showed that the matrix (\ref{m111}), for the pseudoscalar sector, can be written 
as 
\beq
\uppi=\uppi^a \mathsf{t}^a \clt{\sim} \frac{1}{\sqrt 2}
\begin{pmatrix}
\bar{u}\Upgamma u & \bar{d}\Upgamma u & \bar{s}\Upgamma u & \bar{c}\Upgamma u\\
\bar{u}\Upgamma d & \bar{d}\Upgamma d & \bar{s}\Upgamma d & \bar{c}\Upgamma d\\
\bar{u}\Upgamma s & \bar{d}\Upgamma s & \bar{s}\Upgamma s & \bar{c}\Upgamma s\\
\bar{u}\Upgamma c & \bar{d}\Upgamma c & \bar{s}\Upgamma c & \bar{c}\Upgamma c
\end{pmatrix}, \label{p1}
\eeq
evincing the quark-antiquark content of the
mesons in the pseudoscalar channel denoted by $\Upgamma=-i\gamma^{0}\gamma^{1}\gamma^{2}\gamma^{3}$. 	
	For the scalar vacuum expectation value ${\scalebox{.87}{$\upsilon$}}_{\ell,s,c}$, the equations of motion read 
		\begin{eqnarray}
	\left[\partial_z^2 - \left(\frac{3}{z} + \phi'(z)\right)\partial_z
	-\frac{m_5^2}{z^2}\right]{\scalebox{.87}{$\upsilon$}}_q(z)=0, \nonumber
	\end{eqnarray} 
whose solutions are designated by the Tricomi confluent hypergeometric function, $U$, and the associated Laguerre--Sonine polynomial, $L$, as
	\begin{eqnarray}\label{lagu}
		{\scalebox{.87}{$\upsilon$}}_q(z)=\sqrt{\pi}c_1(q)\,z U\left(\frac12,0,\phi\right)-c_2(q)z L\left(-\frac12,-1,\phi\right) \, .
	\end{eqnarray}
	The second term on the right-hand side of Eq. (\ref{lagu}) can be disregarded as long as one intends to acquire a finite action at the IR domain \cite{Colangelo:2008us}. In the UV region, Eq. \eqref{lagu} expands as
	\clt{\begin{eqnarray}
		{\scalebox{.87}{$\upsilon$}}_q(z)\approx 2c_1(q)z+\mu^2\left\{c_2(q)+c_1(q)\left[2\gamma_E-1+2\log(z)+2\log(\mu)+\psi\left(\frac{3}{2}\right)\right]\right\}z^3,
	\end{eqnarray}}
	with the Euler's constant $\gamma_E= \lim _{n\to \infty }\left(\sum _{j=1}^{n}{\frac {1}{j}-\log(n)}\right)\approx 0.577216$ and the digamma function given by $\psi=\frac{d}{dz}\log \Gamma(z)$, where $\Gamma(z)=\int _{0}^{\infty }t^{z-1}e^{-t}{d}t$ is the gamma function.
	The quark mass $m_q$ can be related to the term $c_1(q)$, whereas the quark condensation $\Sigma_q$ can be associated with both functions $c_1(q)$ and $c_2(q)$. The term $c_2(q)$ produces the nonlinear mass spectrum of $a_1$ meson resonances and can be set to $c_2(q) = 0$ \cite{Chen:2021wzj}. One can derive for the auxiliary field $h_c$ in Eq. (\ref{hc}) an analogous expression: 
	\begin{eqnarray}\label{af}
		h_c(z)=D_1\sqrt{\pi}zU\left(\frac{1}{2},0,\phi\right)-D_2L z\left(-\frac{1}{2},-1,\phi\right),
	\end{eqnarray} with the fact that a vanishing value of $D_2$ yields the action to be finite at the IR range of $z$. The difference between the mass spectrum of light-flavor and heavy-flavor vector mesons arises from the mass term of the heavy-flavor quarks involved. 
Comparing (\ref{af}) to the UV behavior of the chiral condensate in the hard-wall model, given by ${\scalebox{.87}{$\upsilon$}}_q(z)= M_q z + \Upsigma_{q} z^3$ \cite{Erlich:2005qh}, one can notice that the quark mass can be expressed by $M_q = 2 c_1(q)$ and the quark condensate $\Upsigma_{q}$ is proportional with $c_2(q)$. 

	The equations of motion of the transverse part of vector fields are obtained when the Euler--Lagrange equations are applied to the action in Eq. (\ref{act3}), resulting in \cite{Chen:2021wzj}
	\begin{eqnarray}\label{eomv}
		\left[\partial_z^2-e^{-\phi(z)}\left(\frac{1}{z^2}+\frac{\phi'(z)}{z}\right)+\frac{2g_5^2}{z^2}\left(\mathsf{m}_V^{ab}-\mathsf{M}_V^{ab}\right)\right] \mathsf{V}^a_{\mu{\scalebox{.54}{$\perp$}}}(q,z)=-q^2\mathsf{V}^a_{\mu{\scalebox{.54}{$\perp$}}}(q,z),
	\end{eqnarray}
	where the $\mathsf{V}^{z,a} = 0$ gauge is taken into account and the  $V_{\scalebox{.54}{$\perp$}}^{\mu,a}$ components satisfy $\partial_\mu V_{\scalebox{.54}{$\perp$}}^{\mu,a}=0$, denoting the Fourier transform by ${\mathsf{V}}^a_{\mu{\scalebox{.54}{$\perp$}}}(z,x)=\frac{1}{(2\pi)^4}\int d^4q e^{iqx}{\mathsf{V}}^a_{\mu{\scalebox{.54}{$\perp$}}}(z,q)$.
AdS/CFT states that the fields $\mathsf{V}^a_{\mu{\scalebox{.54}{$\perp$}}}(q,z)$ can be expressed by $\mathsf{V}^a_{\mu{\scalebox{.54}{$\perp$}}}(q,z)=\mathsf{V}^{(\textsc{n})a}_{\mu{\scalebox{.54}{$\perp$}}}(q,z)+\mathsf{V}^{0a}_{\mu{\scalebox{.54}{$\perp$}}}\mathsf{V}_{{\scalebox{.54}{$\perp$}}}^a(q,z)$, where $\mathsf{V}_{{\scalebox{.54}{$\perp$}}}^a(q,z)$ are bulk-to-boundary 2-point correlators and $\mathsf{V}^{0a}_{\mu{\scalebox{.54}{$\perp$}}}$ emulates the source. 
	The discrete mass spectra $m_n^2=q^2$ and corresponding wave functions can be obtained by taking into account Kaluza--Klein towers of modes  
	$V^{(\textsc{n})a}_{\mu{\scalebox{.54}{$\perp$}}}(q,z)$, with Neumann and Dirichlet boundary conditions $\lim_{z\to0}V^{(\textsc{n})a}_{\mu{\scalebox{.54}{$\perp$}}}(z)=0$ and $\lim_{z\to z_{\textsc{m}}}\left[\partial_zV^{(\textsc{n})a}_{\mu{\scalebox{.54}{$\perp$}}}(z)\right]= 0$, where the index $\textsc{n}$ here labels the Kaluza--Klein excitations. 
	
	\clt{The equations of motion of the transverse part of the axial-vector field, which is defined by $\partial_\mu \mathsf{A}_{\scalebox{.54}{$\perp$}}^{\mu,a}=0$, are expressed as}
	\begin{eqnarray}\label{eoma}
		\left[-\partial_z^2 +  \left(\frac{1}{z} + \phi'(z)\right)\partial_z
+\frac{2g_5^2}{z^2}\mathsf{M}_A^{ab}\right] {\mathsf{A}}^a_{\mu{\scalebox{.54}{$\perp$}}}(z,q)=-q^2{\mathsf{A}}^a_{\mu{\scalebox{.54}{$\perp$}}}(z,q),
	\end{eqnarray}
\clt{where selecting ${\mathsf{A}}^{z,a} = 0$ is a gauge choice}. 
	The boundary conditions of the fields ${\mathsf{A}}^a_{\mu{\scalebox{.54}{$\perp$}}}(z,q)$ and their Kaluza--Klein splitting are identical to the ${\mathsf{V}}^a_{\mu{\scalebox{.54}{$\perp$}}}(z,q)$ fields in the previous paragraph. 
	
	The equations of motion of the longitudinal component of the axial-vector fields and the pseudoscalar fields are written down, respectively, as
		\begin{subequations}
	\begin{eqnarray}	\left[z^2\partial_z^2 - z \left({1} + z{\phi'(z)}\right)  \partial_z\right]\upphi^a(z,q) +
		{2g_5^2\mathsf{M}_A^{ab}}\left[\uppi^a(z,q)-\upphi^a(z,q)\right]&=&0
	\,,\label{eom-pi-1}  \nonumber \\			
			q^2z^2\partial_z \upphi^a(z,q)+{2g_5^2\mathsf{M}_A^{ab}}\partial_z \uppi^a(z,q)&=&0\,.\label{eom-pi-2}  \nonumber
	\end{eqnarray}
	\end{subequations}	
	The boundary conditions  \begin{subequations}\beq
	\lim_{z\to z_{\textsc{m}}}\partial_z\upphi^{(\textsc{n})a}(z,q)&=& 0,\\
	\lim_{z\to0}\uppi^{(\textsc{n})a}(z,q)=\lim_{z\to0}\upphi^{(\textsc{n})a}(z,q)&=&0\eeq\end{subequations}	  are regarded.  
	
	\subsection{Spectroscopy of heavy-light-flavor mesons with a charm quark}
	
The mass spectra of the $D^0$ and $D^*$ heavy-light meson resonances with charm, and the charmonium-like $\eta_c$  and $\chi_{c1}$ families, can be obtained by appropriately choosing the physical parameters in the 4-flavor AdS/QCD that best fit experimental data in PDG \cite{pdg}. The  coefficient $\mu$ entering the quadratic dilaton (\ref{dila}), the IR hard-wall cutoff $z_{\textsc{m}}$, and the VEV of $c_1(q)$, for the four quark flavors $q=(\ell, s, c)$, are fixed by the experimental values of the masses of the mesons 
${\scalebox{.985}{$\psi$}}(3770)$,
 $J/{\scalebox{.985}{$\psi$}}(1S)$, 
 $K^{*}(892)$, 
 ${\scalebox{.985}{$\chi_{c1}$}}(3872)$,  
 $a_{1}(1260)$, and 
 $\rho(770)$. They are presented in Table \ref{fixed}. 
\begin{table}[H]
\begin{center}\medbreak
\begin{tabular}{||c|c|c|c|c|c||}
\hline\hline
 State & Constituent quarks& $I^G\left(J^{PC}\right)$&$\;M_{\scalebox{.64}{\textsc{Exp.}}}$ (MeV)\;  & $M_{\scalebox{.64}{\textsc{AdS/QCD}}}$ (MeV) &\;RE (\%)\;\\
       \hline\Xhline{2\arrayrulewidth}
\;$\rho(770)\;$ &$\bar{u}u/\bar{d}d$ & $1^+(1^{--})$& $775.26\pm0.23$ & $860.66$  & 9.85 \\ \hline
\;$K^{*}(892)\;$ & $\bar{d}s/\bar{s}d$&$\frac12(1^-)$& $891.67\pm 0.26 $ & $884.37$ & 0.86 \\ \hline
$J/{\scalebox{.985}{$\psi$}}(1S)$& $\bar{c}c$&$0^-(1^{--})$ &\;$3096.900 \pm 0.006$\;       &\; $3098.112$\; &  0.03   \\\hline
\;$a_{1}(1260)$\;& $\bar{u}d$ &$1^-(1^{++}) $&$1230\pm 40$& 1222&0.65 \\\hline
${\scalebox{.985}{$\chi_{c1}$}}(1P)$&$\bar{c}c$ & $0^+(1^{++})$ & $3510.67\pm 0.05$&$3464.12$&1.34 \\\hline
${\scalebox{.985}{$\psi$}}(3770)$& $\bar{c}c$&$0^-(1^{--}) $ &$3778.1\pm 0.7$&$3712.1$&0.16\\
\hline\hline
\end{tabular}
\caption{Mass spectrum of the $\rho(770)$, $K^{*}(892)$, $J/{\scalebox{.985}{$\psi$}}(1S)$, $a_1(1260)$, ${\scalebox{.985}{$\psi$}}(3770)$,  ${\scalebox{.985}{$\chi_{c1}$}}(1P)$ meson state resonances: experimental data (fourth column) and 4-flavor AdS/QCD  (fifth column).  The sixth column displays the relative error per state. } \label{fixed}
\end{center}
\end{table}
 The value  $\mu\approx 0.432$ GeV is chosen to match the  Regge trajectory slope of the $\rho$ meson family and their experimental mass. Moreover, the value of the hard-wall cutoff $z_{\textsc{m}}$ can be attained by the charmonia $J/{\scalebox{.985}{$\psi$}}$ and ${\scalebox{.985}{$\psi$}}(3770)$  masses, and their   Regge trajectory slope as well. Eq. (\ref{eqk}) and the masses and Regge slopes for the $a_1^-$ and $K^{*-}$ meson resonances are employed to fix the values $c_1(\ell)$, $c_1(s)$, and $c_1(c)$ which can also be read off, through the expansion of
the VEV of ${\scalebox{.87}{$\upsilon$}}_{\ell, s,c}$, the quark masses $M_{\ell,s,c}$ and the quark condensation $\Upsigma_{\ell,s,c}$.  The masses of the  strange and charm quarks are fitted to the experimental  masses for the mesons.   To accomplish it, one fixes $M_{\ell}\sim140.2~$MeV, $M_{s}\sim200$ MeV, $M_{c}\sim1200$ MeV, $\Upsigma_{\ell}= 2.4601\times 10^6~\rm{MeV}^{3}$, $\Upsigma_{s}=3.5119\times 10^6~\rm{MeV}^{3}$,  $\Upsigma_{s}=2.1023\times 10^7~\rm{MeV}^{3}$,   $M_c=1020.5~$MeV and $\Upsigma_{c}=1.7983\times 10^7~\rm{MeV}^{3}$  \cite{Chen:2021wzj}.
	The complete set of predictions derived from the 4-flavor AdS/QCD model, together with the experimental data of the meson resonances for in the $D^0$, $D^*$, $\eta_c$, and $\chi_{c1}$  families in PDG \cite{pdg}, are respectively displayed and discussed in Tables \ref{scalarmasses1},  \ref{scalarmasses2}, \ref{scalarmasses3}, and \ref{scalarmasses4}. 

\begin{table}[H]
\begin{center}\medbreak
\begin{tabular}{||c||c|c|c|c||}
\hline\hline
\;$n$\;& State ($\bar{c}u/\bar{u}c$) &$M_{\scalebox{.64}{\textsc{Exp.}}}$ (MeV)  & $M_{\scalebox{.64}{\textsc{AdS/QCD}}}$ (MeV) &\;RE (\%)\;\\
 \hline\Xhline{2\arrayrulewidth}
1 &\;$D^0\;$ &\;$1864.85\pm0.05$\; & $1671.14$  & 10.31 \\ \hline
2 &\;$D(2550)^0\;$ & $2549\pm19 $ & $2778$ & 8.24 \\ \hline
3 &\;$D(3000)^0\;$ &\; $3214\pm29\pm49 $\; & $3317$ & 3.10 \\ 
\hline\hline
\end{tabular}
\caption{Mass spectrum of the neutral $D^0$ meson resonances:   experimental values and AdS/QCD model with four flavors including the charm quark.  The fifth column displays the relative error per state. } \label{scalarmasses1}
\end{center}
\end{table}
\clt{It is worth emphasizing that in Table \ref{scalarmasses1}, the $D(3000)^0$ resonance  is displayed with experimental mass $(3214\pm29\pm49)$ MeV, which is extracted from the PDG \cite{pdg}. Explicitly, the mass of the $D(3000)^0$ resonance is given by $(3214\pm29_{\rm stat} \pm49_{\rm syst})$ MeV, where the first error refers to the statistical error, whereas the second error refers to the systematic one. 
This notation, used by the PDG, is fixed hereon throughout the text.}

The Regge trajectory profile is a relevant tool for studying meson spectroscopy, additionally probing the meson inner structure and  imparting immediate clues about the quark flavors constituting the mesons. 
The linear Regge trajectory, relating the squared mass spectrum of the $D^0$ meson family as a function of the principal quantum number $n$, fits the mass spectrum in Table \ref{scalarmasses1}, regarding experimental data reported in PDG \cite{pdg}, reading   
\beq 
m_n^2 =3.426 \,n-0.083,\label{lrt0}
\eeq
within 2.25\% RMSE (root mean square error). When the dilaton is, as usual, a quadratic function of the energy scale in the AdS${}_5$ bulk, it  implies the Regge trajectory to be linear, $m_n^2\propto n$, which in addition prevents any ambiguity of choice about the IR boundary conditions \cite{Karch:2006pv}.  

 It is worth pointing out that the neutral $D^0$ meson resonance comprises the lightest heavy-light meson  containing a single charm quark [antiquark] and has to change the charm [anti]quark into another kind of [anti]quark to decay, which 
can only occur through the weak force interaction. The quantum numbers for the $D^0$ meson resonances, as well as for other excited charmed mesons were reported in Ref. \cite{LHCb:2019juy}. In the $D^0$ heavy-light meson family, the charm quark preferentially changes into a strange quark via an exchange of a $W$ boson. Hence, the $D^0$ meson family decays preferentially into kaons and pions, although muons are observed, rarely though  \cite{pdg}.

Next,  the mass spectrum of the neutral $D^*$ meson resonances constituted by heavy-light $\bar{u}c$ bound quarks, with $I(J^P)=\frac12(1^-)$,  is displayed and discussed.  
\begin{table}[H]
\begin{center}\medbreak
\begin{tabular}{||c||c|c|c|c||}
\hline\hline
\;$n$\;& State ($\bar{c}u/\bar{u}c$) &$M_{\scalebox{.64}{\textsc{Exp.}}}$ (MeV)  & $M_{\scalebox{.64}{\textsc{AdS/QCD}}}$ (MeV) &\;RE (\%)\;\\
 \hline\Xhline{2\arrayrulewidth}
1 &\;$D^*(2007)^0$\; &\;$2006.85\pm0.05$\; & $2296.33$  & 12.63 \\ \hline
2 &\;$D^*_1(2600)^0\;$ & $2627\pm10 $ & $2512$ & 4.37 \\ \hline
3 &\;$D^*_1(2760)^0\;$ &\; $2781\pm18\pm13 $\; & $2756$ & 0.89 \\ 
\hline\hline
\end{tabular}
\caption{Mass spectrum of the neutral $D^*$ meson resonances:   experimental values and AdS/QCD model with four flavors including the charm quark.  The fifth column displays the relative error per state.  } \label{scalarmasses2}
\end{center}
\end{table}
\noindent Table \ref{scalarmasses2} displays the mass spectrum of the already observed heavy-light-flavor charmed resonances $D^*(2007)^0$, $D^*_1(2600)^0$, and $D^*_1(2760)^0$.  
The BaBar analysis of the angular distribution points to identifying 
 the $D^*_1(2600)^0$ meson to the first radial excitation of an $S$-wave charmed meson, whereas the $D^*_1(2760)^0$ can be a
 matched to a $D$-wave open charm state \cite{Chen:2016spr,pdg}.

In both $D$-meson families in Tables \ref{scalarmasses1} and \ref{scalarmasses2} already discussed, the  $D(2550)^0$, $D^*_1(2600)$, and $D^*_1(2760)$ meson states were for the first time experimentally detected in inclusive electron-positron collisions by BaBar collaboration, mainly within the charmed channel 
$e^+e^-\to c\bar c\to D^{\star}\pi Y,$
where $Y$ is any collision byproduct~\cite{BaBar:2010zpy}. Also, the $D(2550)^0$ and $D^*_1(2600)$ meson resonances  consist of a concrete realization of the predicted radial excitations $2S$, $^1S_0$, and $^3S_1$, respectively\footnote{In the terminology of hadronic spectroscopy, the $n, S, J$ quantum numbers together with the magnetic quantum number $M_J$ label a specific state, with standard symbol ${}^{2S+1}n_J$. The quantum number $n$ uses the spectroscopic notation. The letter ``$S$'' corresponds to $n=0$, the letter ``$P$'' denotes $n = 1$, and the letter ``$D$'' stands for $n=2$, whereas ``F''  represents $n= 3$, and so on.}. Regarding $pp$ and $e^+e^-$ collisions, one observes the inclusive production of the natural parity 
resonances $D^*_1(2600)^0$ and $D^*_1(2760)^0$, in $D^{*+}\pi^-$ channels. 
In addition, highly excited heavy-light-flavor meson resonances can be yielded in exclusive $B$  decays, involving meson resonances constituted by a bottom antiquark and either an up ($B^+$), down ($B^0$), strange ($B^0_s$), or  
 charm quark ($B^+_c$). In the decay channel $B^-\to D^+K^-\pi^-$, the charmed $D^*_1(2760)$ meson resonance can be observed 
 \cite{LHCb:2015eqv,LHCb:2015klp}. 
 
Hadronic decays of heavy-light mesons produce valuable information about their quark dynamics. Since the experimental detection of the $D(2550)^0$, $D^*_1(2600)$, and $D^*_1(2760)$ meson resonances, their hadronic decays have been examined in several theoretical approaches. 
In a chiral quark $^3P_0$ model, which assumes a creation of a $q\bar{q}$ pair from the
vacuum with the corresponding quantum numbers $J = 0$, $n = 1$, and $S =1$,
 the $D^*_1(2600)$ meson resonance can be realized as an admixture of the radially excited  $2^3S_1$ and the orbitally excited $1^3D_1$ meson states, with $J^P=1^-$ \cite{Bernardini:2003ht}. The 
 $D^*_1(2760)$ meson can be identified as either the 
orthogonal partner of the $D^*_1(2600)$ charmed meson or the $1^3D_3$ state, whereas the $D(2550)^0$ meson resonance has a decay width consistent with the experimental value of a  $2^1S_0$ state, in a heavy-flavor quark effective theory  \cite{Zhao:2016mxc}. Moreover, in the $^3P_0$ model, the $D(2550)^0$ and the $D^*_1(2600)$ resonances are regarded as the  $2^1S_0$ and $2^3S_1$ states, respectively. On the other hand, the $D_1^*(2760)$ can be realized as an admixture of the $1^1D_2$ and $1^3D_2$ states ~\cite{Zhong:2010vq,Ni:2021pce,Sun:2010pg,Wang:2010ydc,Chen:2011rr,Ge:2015fxa}.

The linear Regge trajectory, relating the squared mass spectrum of the $D^*$ meson family as a function of the principal quantum number $n$, fits experimental data in PDG \cite{pdg}, and reads  
\beq 
m_n^2 =  1.853\,n+2.514,\label{lrt}
\eeq
within 5.49\% RMSE. 
The Regge trajectory (\ref{lrt})  for the $D^*$ heavy-light-flavor mesons is linear with relatively good accuracy, with the slope $\alpha\approx 1.85$ GeV${}^{2}$. 

\subsection{Spectroscopy of charmonium-like meson resonances}

The dissociation temperature of quarkonium states depends on their binding energy, with strongly bound states like 
${\displaystyle J/\psi }$ and 
$\upsilon(1S)$ melting at higher temperatures compared to loosely bound states such as 
${\displaystyle \psi (2S)}$ and ${\displaystyle \chi _{c1}}$ for the charmonium family. The dissociation process permits to employ the charmonium dissociation probabilities to determine the medium temperature, assuming that quarkonium dissociation is the main mechanism involved \cite{Bernardini:2003ht,Krein:2017usp,Cobos-Martinez:2020ynh,Yin:2019bxe}. Therefore, investigating the mass spectrum of the next generation of charmonia in the families $\eta_c$ and $\chi_{c1}$ can provide additional physical aspects of charmonia. Firstly, the $\eta_c$ charmonium-like meson family, with $I^G(J^{PC}) = 0^+(0^{-+})$ quantum numbers, can be addressed. 
\begin{table}[H]
\begin{center}\medbreak
\begin{tabular}{||c|c||c|c|c||}
\hline\hline
$n$ & State ($\bar{c}c$) & $M_{\scalebox{.64}{\textsc{Exp.}}}$ (MeV)  & $M_{\scalebox{.64}{\textsc{AdS/QCD}}}$ (MeV) &\;RE (\%)\;\\
       \hline\Xhline{2\arrayrulewidth}
1 &\;$\eta_c(1S)\;$ & $2984.1\pm0.4$ & $2600.8$  & 12.87 \\ \hline
2 &\;$\eta_c(2S)\;$ & $3637.7\pm0.9 $ & $3641.2$ & 0.09 \\ \hline\hline
\end{tabular}
\caption{Mass spectrum of the $\eta_c$ charmonium-like meson:   experimental values and AdS/QCD model with four flavors including the charm quark.  The fifth column displays the relative error per state. } \label{scalarmasses3}
\end{center}
\end{table}
\noindent  The $\eta$ and $\eta^\prime$ are well known to be isosinglet mesons made of a mixture of up/down, and strange quarks, and their antiquarks. The  $\eta_c$ charmonium-like meson family consists of similar forms of  quarkonium, as they have the same spin and parity as the light-flavor $\eta$ family. However, the $\eta_c$ charmonium-like meson has only $\bar{c}c$ bound quarks.  
Heavy quarkonium can be explored in effective field
theoretical approaches, taking into account the charmonia $J/\psi$ and $\eta_c$ charmonium-like production from $e^+e^-$ scatterings. Similar analyses can be extended to the charmonium in the higher resonances like $\psi(2S)$, $\eta_c(2S)$, and $\chi_{c1}$ as well  \cite{Yu:2016pqp,Barbosa-Cendejas:2018mng,Ferreira:2019nkz}.


The $\chi_{c1}$ charmonium-like meson family, with $I^G(J^{PC}) = 0^+(1^{++})$, can be addressed. 
\begin{table}[H]
\begin{center}\medbreak
\begin{tabular}{||c|c||c|c|c||}
\hline\hline
$n$ & State ($\bar{c}c$) & $M_{\scalebox{.64}{\textsc{Exp.}}}$ (MeV)  & $M_{\scalebox{.64}{\textsc{AdS/QCD}}}$ (MeV) &\;RE (\%)\;\\
       \hline\Xhline{2\arrayrulewidth}
1 &\;$\chi_{c1}(1P)\;$ & \;$\;3510.67\pm0.05$\; & $3464.72$  & 1.33 \\ \hline
2 &\;$\chi_{c1}(3872)\;$ & \;$3871.64\pm0.06$\; & $3808.17$  & 1.67 \\ \hline
3 &\;$\chi_{c1}(4140)\;$ & $4146.5\pm3.0$ & $4138.1$  & 0.20 \\ \hline
4 &\;$\chi_{c1}(4274)\;$ & $4286^{+8}_{-9}$ & $4460$  & 3.90 \\ \hline
5 &\;$\chi_{c1}(4685)\;$ & $4684\pm7^{+13}_{-16}$ & $4723$  & 0.82 \\
\hline\hline
\end{tabular}
\caption{Mass spectrum of the neutral $\chi_{c1}$ charmonium-like meson   resonances:   experimental values and AdS/QCD model with four flavors including the charm quark.  The fifth column displays the relative error per state. } \label{scalarmasses4}
\end{center}
\end{table}
\noindent Some phenomenological challenge sets in when the meson $
\chi_{c1}(3872)$ is taken into account, as its mass coincides with the sum of the masses of the $D^0$ and $D^{*0}$ mesons \cite{pdg}.  Also, evidence of $\chi_{c1}(3872)$  production in the quark-gluon plasma was reported \cite{CMS:2021znk}. 
The linear Regge trajectory, relating the squared experimental mass spectrum of $\chi_{c1}$ as a function of the principal quantum number $n$ reads   
\beq 
m_n^2 &=&2.261\,n+10.18,\label{lrt2}
\eeq
within a 2.8\% RMSE.

\section{DCE of $D^0$, $D^*$, $\eta_c$, and $\chi_{c1}$ meson resonances and mass spectrum of the next generation of heavy-light mesons with charm and charmonia}
\label{sec2}

One can compute the DCE of the meson families reported and discussed in Sec. \ref{sec1} as long as the energy density $\rho(X^M)$, namely, the timelike component of the stress-energy-momentum tensor underlying the 4-flavor AdS/QCD model, is taken into account to represent the physical system in QCD. The energy density can be calculated by the expression
\begin{flalign}\label{rhom}
  \rho(X^M)=  \tau_{00}(X^M) = \frac{2}{\sqrt{-g}}\left[ \frac{\partial(\sqrt{-g}\mathcal{L})}{\partial g^{00}} - \frac{\partial}{\partial x^N}\frac{\partial(\sqrt{-g}\mathcal{L})}{\partial\left(\frac{\partial g^{00}}{\partial x^N}\right)}
    \right],
\end{flalign}
where $g$ denotes the determinant of the metric (\ref{eq.1}) and $\mathcal{L}$ denotes the Lagrangian density, which can be read off the action (\ref{act1}) governing the 4-flavor AdS/QCD model, whose quartic order expansion consists of the integrand of the actions (\ref{act2}) - (\ref{act5}). 
The first step used to compute the DCE consists of considering the energy density in momentum space, by taking a Fourier transform of Eq. (\ref{rhom}), which is localized and Lebesgue-integrable, as 
\begin{flalign}\label{fou}
    \tau_{00}(q_M) = \frac{1}{(2\pi)^{M/2}} 
    \int_{-\infty}^{+\infty}\ldots\int_{-\infty}^{+\infty}
    \tau_{00}(x^M)e^{-iq_N x^N}\,{\sqrt{-g}}\,d^Mx,
\end{flalign}
where $q_M = (q_\mu, q_z)$ is the 5-momentum associated with the components $x^M = (x^\mu,z)$ of the position vector. The symbolic integration in Eq. (\ref{fou}) 
is, a priori, taken over the whole AdS bulk space, although we will see promptly after Eq. (\ref{dce}) that a more refined analysis can simplify our integrations. 
 Subsequently, the protocol to compute the DCE asserts that the modal fraction encodes the power spectral density and can be expressed by 
\cite{Gleiser:2018kbq}, 
\begin{flalign}
    \boldsymbol{\tau}_{00}(q_N) = \frac{\abs{\tau_{00}(q_N)}^2}{\displaystyle\int_{-\infty}^{+\infty} \ldots\int_{-\infty}^{+\infty}\abs{\tau_{00}(q_M)}^2 \,d^M{q}}.
    \label{dce_modal}
\end{flalign} It corresponds to the weight of individual ${q}$ wave modes and their contribution to the whole energy density profile portraying the QCD system. 
The modal fraction (\ref{dce_modal}) encodes the Fourier transform of the 2-point field correlator function of the fluctuations in the energy density and represents the whole set of modes with momentum $q_M$ that promotes the localized shape of the energy density.

The DCE, hence, encodes the quantity of information involved to describe the profile of the energy density, as 
\begin{flalign}
    \text{DCE} = -\int_{-\infty}^{+\infty} \ldots\int_{-\infty}^{+\infty} \check{\boldsymbol{\tau}}_{00} (q_N) \log\check{\boldsymbol{\tau}}_{00}(q_N) d^Nq,
    \label{dce}
\end{flalign}
where \beq\label{chec}
\check{\boldsymbol{\tau}}_{00}(q_N)=\frac{\boldsymbol{\tau}_{00}(q_N)}{\boldsymbol{\tau}_{00}^{\mathsf{sup}}(q_N)},\eeq for ${\boldsymbol{\tau}}_{00}^{\mathsf{sup}}(q_N)$ standing for the supremum value of the energy density $\boldsymbol{\tau}_{00}(q_N)$, taking into account the entire integration range. 
The DCE supports the  link between dynamical and informational contents of physical systems in QCD with localized energy configurations, as is the case of charmed mesons here studied.

The index $M=1$, associated with $X^5\equiv z$,   will be fixed in Eq. (\ref{fou}), whereas $N=1$ will be regarded in Eqs. (\ref{dce_modal}) and (\ref{dce}) to estimate the DCE, since a Kaluza--Klein splitting has been taken into account already. In fact, the AdS boundary has codimension one with respect to the AdS bulk, in the 4-flavor AdS/QCD, and the $z$ coordinate is the energy scale of QCD (see, e.g.,  Refs. \cite{Aharony:1999ti,Pirner:2009gr}). \clt{The DCE 
can be evaluated in terms of the energy scale, taking into account all the possible values of $z$. The AdS boundary has codimension one with respect to the AdS bulk in the 4-flavor AdS/QCD. Therefore, the energy density is computed along the bulk coordinate $z$,  as the dilaton, the warp factor,  and indeed all the functions in the action (\ref{act2}) - (\ref{act5}) are $z$-dependent only, as can be realized by Eqs. (\ref{x0}) and Eqs. (\ref{t11}) - (\ref{t15}). It makes the energy density vanish at the AdS boundary. Therefore Eqs. (\ref{fou}) - (\ref{dce}) can be integrated out of the AdS boundary, across the AdS bulk. }

For computing the DCE, numerical integration can be implemented and  Eq. (\ref{fou}) can be expressed, after 
a change of variables from $z\in[0,+\infty)$ to ${\scalebox{.93}{$\mathfrak{z}$}}\in[0,1]$, as an integral over a finite interval $[0,1]$, instead of taking the entire real line as the integration interval: 
\begin{eqnarray}\label{fou1}
    \tau_{00}(q) &=& \frac{1}{\sqrt{2\pi}} \int_{0}^{+\infty}\tau_{00}(z)e^{-iqz}dz\nonumber\\ &=& \frac{1}{\sqrt{2\pi}}\int _{0}^{1}\tau_{00}\left({\frac{{\scalebox{.93}{$\mathfrak{z}$}}}{1-{\scalebox{.93}{$\mathfrak{z}$}}}}\right)
    \exp\left[-iq\left({\frac{{\scalebox{.93}{$\mathfrak{z}$}}}{1-{\scalebox{.93}{$\mathfrak{z}$}}}}\right)
   \right]\frac{1}{\left(1-{\scalebox{.93}{$\mathfrak{z}$}}\right)^{2}}\,d{\scalebox{.93}{$\mathfrak{z}$}}.
\end{eqnarray}
It is worth pointing out that the integrals in Eqs. (\ref{fou}) - (\ref{dce}) are computed over the real line $(-\infty, +\infty)$. Nevertheless, as the AdS coordinate $z$ is essentially the QCD energy scale, the integration range consists, for practical reasons, of the non-negative 
 real axis $z\in[0,+\infty)$. It is worth emphasizing that the fluctuation wave functions solved from the fluctuation equations for each mode are inserted one by one, setting all other fields to zero. This follows an analogous procedures of Refs.~\cite{Braga:2018fyc,MarinhoRodrigues:2020ssq}. 
\clt{In fact, the radial wave functions for the fluctuations are inserted, and the plane-wave terms are dropped out. One requires that the gauge fields be real when plugging in the plane waves. } 
 
 Analogously, the integration over the real line in Eq. (\ref{dce}) can be rewritten within a finite integration interval as 
\begin{flalign}
    \text{DCE} = -\int_0^1 \check{\boldsymbol{\tau}}_{00}\left({\frac{{\scalebox{.93}{$\mathfrak{z}$}}}{1-{\scalebox{.93}{$\mathfrak{z}$}}}}\right)
    \log\left\{\check{\boldsymbol{\tau}}_{00}\left({\frac{{\scalebox{.93}{$\mathfrak{z}$}}}{1-{\scalebox{.93}{$\mathfrak{z}$}}}}\right)\right\} 
    \frac{1}{\left(1-{\scalebox{.93}{$\mathfrak{z}$}}\right)^{2}}\,d{\scalebox{.93}{$\mathfrak{z}$}}.
    \label{dce1}
\end{flalign}
Eq. (\ref{dce1}) is a general expression that holds for arbitrary cases. 

The DCE is solved by the Newton--Cotes quadrature method, using the composite iterated trapezoidal rule. The output convergence is observed as long as the grid partition is progressively refined. For $16384 \;(= 2^{14})$ grid points, the numerical error is kept below $10^{-7}$, for each DCE value (\ref{dce1}) taking between 7.2 and 8.0 minutes on an 8 Core 4.8 GHz i9, under OsX Sequoia. For completeness,  the 
 Newton--Cotes quadrature method is also implemented 
using Simpson's rule, giving the same convergent numerical results with the numerical error within $10^{-8}$. For completeness,  Boole's rule provides the numerical error within $10^{-9}$. These two last computational refinements of the composite iterated trapezoidal rule demonstrate the robustness of the exhaustive numerical method employed here to compute the DCE.   Since the DCE (\ref{dce1}) is  computed  for 13 charmed meson resonances in the $D^0$, $D^*$,  $\eta_c$, and $\chi_{c1}$  families, in the next subsections, the numerical analysis, although repetitive, is straightforwardly feasible.

With the calculated DCE values, it is possible to express them both as a function of the principal quantum number $n$ and as a function of the squared mass $m^2$ of each charmed mesonic state. Each one of the heavy-light-flavor $D^0$ and $D^*$ meson families has three states already detected in PDG, whereas the $\eta_c$ charmonium family has two  detected states, and the $\chi_{c1}$ has five resonances reported in PDG \cite{pdg}. 
Let us denote by $n_0$ the number of charmed meson states reported in PDG,  in each of the meson families here to be analyzed. Polynomial interpolation of the DCE as a function of $n$ naturally permits attributing new meson states labeled by  $n > n_0$. For the $D^0$, $D^*$, and $\eta_c$ families, new heavier resonances in the respective next generations will be studied hereon by considering $n=n_0+1$ and $n=n_0+2$. For the $\chi_{c1}$ charmonium family, since there are five of these resonances reported in PDG, for $n=1,\ldots,5$, then the new charmonia states corresponding to $n=6,7,8$ will be scrutinized. Irrespective of which meson family 
will be regarded, the first kind of DCE-Regge-like trajectory depicts the DCE as a function of $n$. It is therefore feasible to also take into account the DCE as a function of the experimental mass spectrum in each of the $D^0$, $D^*$,  $\eta_c$, and $\chi_{c1}$ charmed meson families and 
extrapolate them to estimate the mass spectrum of a new generation of heavier charmed meson resonances, in each one of these families. These higher-excited charmed mesonic resonances may be identified as meson resonances that have not been matched to further meson states omitted from the summary table in PDG, yet. It is one of the leading phenomenological features of employing the DCE in the context of 4-flavor AdS/QCD.
Besides the power to reckon the mass spectrum of the next generation of charmed meson resonances, the DCE approach circumvents a caveat in AdS/QCD regarding meson spectroscopy, allowing for the estimation of the mass spectrum of heavier charmed meson resonances from interpolation of experimental data. It takes into account the mass spectrum of charmed mesons already detected and reported in PDG \cite{pdg}. From the phenomenological point of view, 
this method is more concise and accurate than pure AdS/QCD, which obtains the mass spectrum as the squared eigenvalue of the Schrödinger-like operators, which come from the Euler--Lagrange equations governing the meson resonances in each $D^0$, $D^*$, $\eta_c$, and $\chi_{c1}$ meson  families. 

The power spectral density encoded by the modal fraction and the DCE underlying the 4-flavor AdS/QCD are computed for the charmed meson families following the protocol given by Eqs. (\ref{fou}, \ref{dce_modal}, \ref{dce}) or, equivalently, Eqs. (\ref{fou1}, \ref{dce1}). Therefore, the mass spectrum of the next generation of charmed mesons in the  $D^0$, $D^*$, $\eta_c$, and $\chi_{c1}$ families can be then yielded by direct interpolation of the experimental mass spectra of the resonances in each one of these four families in PDG \cite{pdg}. The  $D^0$ charmed meson family will be the first family to be addressed in the next subsection.

 \subsection{DCE of charmed $D^0$ meson resonances}
  \label{ndce}

\begin{figure}[H]
	\centering
	\includegraphics[width=9.8cm]{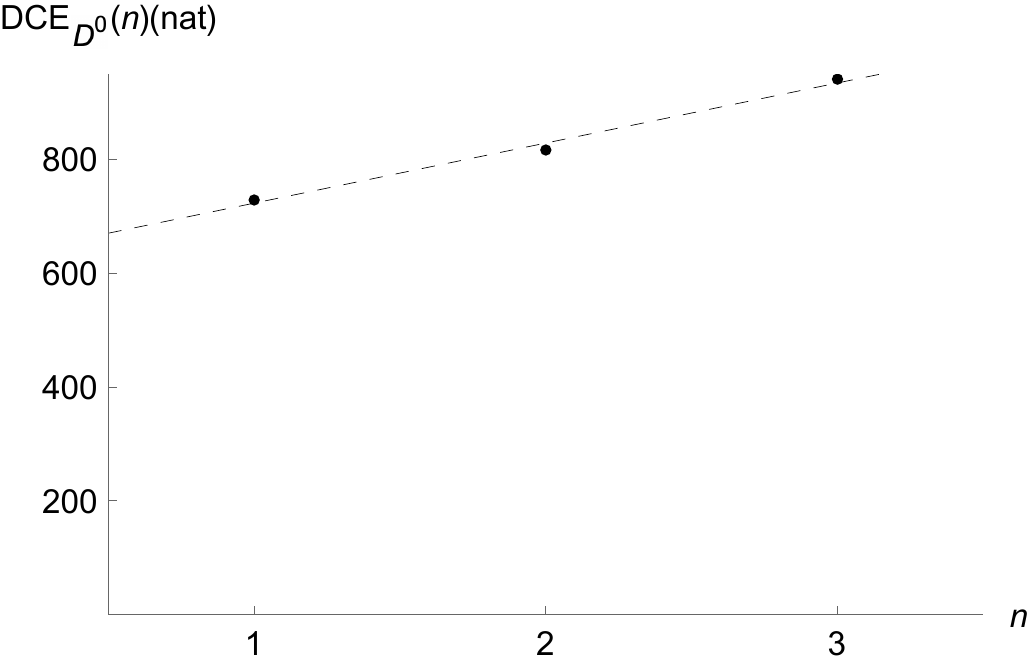}
	\caption{DCE of $D^0$ meson resonances as a function of the principal quantum number $n$, for  $n=1,2, 3$ (respectively corresponding to the states $D^0$,  $D^0(2550)^0$, and $D^0(3000)^0$ \cite{pdg}).  
The DCE-Regge trajectory of the first  kind, Eq. (\ref{itp1}), is plotted by a dashed line.}
	\label{cen1d}
\end{figure}

 We start by calculating the DCE of the $D^0$ meson family. The first  resonance, $n=1$,  corresponds to the 
$D^0$ meson. Taking $n=2$, the state $D^0(2550)^0$ is described, whereas for $n=3$ one acquires the $D^0(3000)^0$ charmed meson resonance.  
The DCE  can then be evaluated, using the energy density \eqref{rhom} into Eqs. (\ref{fou}, \ref{dce_modal}, \ref{dce}) or, equivalently, Eqs. (\ref{fou1}, \ref{dce1}). The DCE method underlying 4-flavor AdS/QCD is a realistic setup that emulates the estimates about the meson mass spectra of bottom-up AdS/QCD, providing a robust and trustworthy approach to obtain the mass spectrum of the next generation of charmed meson resonances. In particular, it will take into account the experimental values of the mass spectrum of the $D^0$,  $D^0(2550)^0$, and the $D^0(3000)^0$  to obtain the mass spectrum of the next generation of  $D^0$ meson resonances. The values of the DCE, numerically computed for the $D^0$ meson family, are listed in Table \ref{scalarmasses50}.
\begin{table}[H]
\begin{center}
\begin{tabular}{||c|c|c||}
\hline\hline
$n$ & \;{} $D^0$ meson resonances\; & DCE (nat) \\
       \hline\Xhline{2\arrayrulewidth}
\;1\; &\;$D^0\;$ & 728.65   \\ \hline
\;2\; &\;$D^0(2550)^0\;$ & 817.89   \\ \hline
3 &\;$D(3000)^0\;$ & $939.80$  \\ 
\hline\hline
\end{tabular}
\caption{DCE of the $D^0$ meson family, for $n=1, 2, 3$.} \label{scalarmasses50}
\end{center}
\end{table}
\noindent When the DCE is expressed as a function of the $n$ principal quantum number, linear  interpolation of data in Table \ref{scalarmasses50} yields the  DCE-Regge trajectory of the first  kind, 
\begin{eqnarray}\label{itp1}
\!\!\!\!\!\!\!\!\!\!\!\!{{\rm DCE}}_{D^0}(n)&=& 105.575\,n+617.630,  \end{eqnarray}
within $0.012\%$ RMSE. 
Data in Table \ref{scalarmasses50} and Eq. (\ref{itp1}) are together depicted in Fig. \ref{cen1d}.

\begin{figure}[H]
	\centering
	\includegraphics[width=9.8cm]{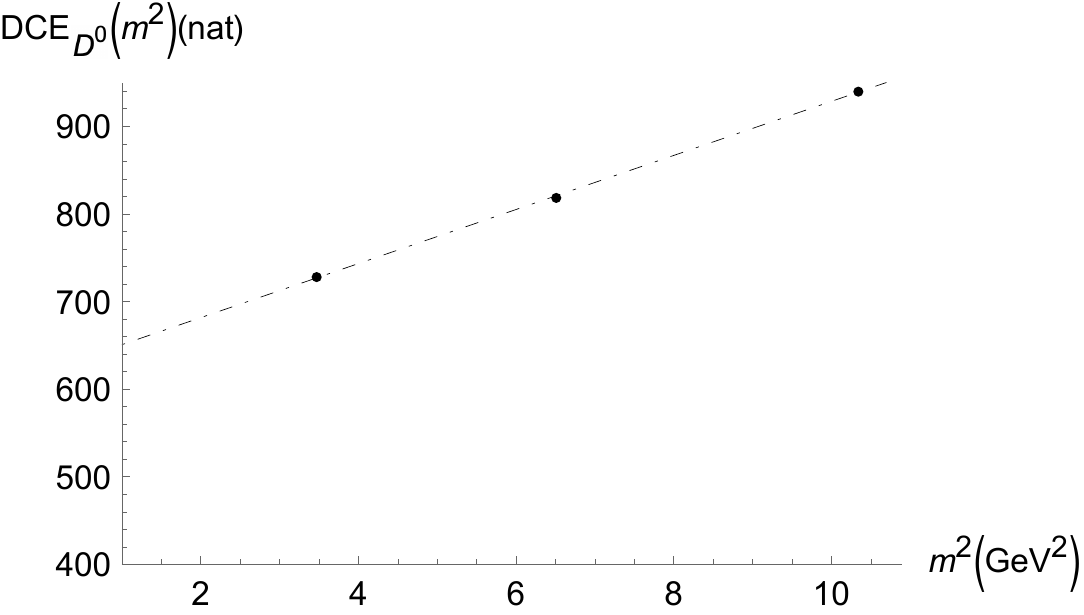}
	\caption{DCE of the $D^0$ meson family displayed as a function of their respective squared mass, for  $n=1,2,3$ (corresponding, respectively, to the $D^0$,  $D^0(2550)^0$, and $D^0(3000)^0$ heavy-light-flavor meson states \cite{pdg}). 
The DCE-Regge trajectory of the second kind in Eq. (\ref{itq11}) corresponds to the interpolating dot-dashed line.}
	\label{dcedom2}
\end{figure}

In addition, the DCE underlying the $D^0$ charmed meson family can be arranged with respect to the experimental mass spectrum of the $D^0$ meson states. Hence, the DCE of the $D^0$ meson family, listed in Table \ref{scalarmasses50},  can be plotted, instead, as a function of the squared mass of each $D^0$ meson state, presented in Table \ref{scalarmasses1}. The resulting data are displayed in Fig. \ref{dcedom2}, with an interpolation curve corresponding to a DCE-Regge trajectory of the second kind, given by Eq. (\ref{itq11}). 
\noindent The DCE-Regge trajectory of the second kind, expressing the DCE of the $D^0$ meson family with respect to their respective (squared) mass spectrum, $m^2$ (GeV${}^2$), is given by the interpolation 
\begin{eqnarray}
\label{itq11}
\!\!\!\!\!\!\!\!\!\!\!{\rm DCE}_{D^0}(m^2) \!\!&\!=\!& 30.859\, m^2 + 619.917.
   \end{eqnarray} within $0.01\%$  RMSE.    
   
Eqs. (\ref{itp1}) and (\ref{itq11}) are two kinds of DCE-Regge trajectories, containing the features needed to deploy the mass spectrum of the next generation of meson resonances within the $D^0$ family, corresponding to values $n>3$. When these values are substituted in Eq. (\ref{itp1}), the respective values of the DCE can be read off and replaced onto the left-hand side of Eq. (\ref{itq11}), which can be algebraically solved. Therefore, this process engenders the mass spectrum of each resonance in the next generation in the $D^0$ family. This technique relies on the DCE, underlying the 4-flavor AdS/QCD, and on the experimental mass spectrum of the $D^0$,  $D^0(2550)^0$, and $D^0(3000)^0$ heavy-light-flavor meson states in PDG \cite{pdg}. Hence, this protocol is more realistic, from the phenomenological point of view, than the method in 4-flavor AdS/QCD of reading off the mass spectrum of $D^0$ resonances as eigenvalues of the Schr\"odinger-like equations. 
The DCE-Regge trajectory of the second  kind, relating the DCE of the $D^0$ meson family as a function of the squared mass spectrum, is 
constructed when the experimental mass spectrum of the $D^0$ meson states is interpolated. This procedure is shown in Fig. \ref{dcedom2}.

To get the  mass spectrum of the next generation of charmed meson resonances in the $D^0$ meson family, for $n>3$, one must start with the case $n=4$, related to the $(D^0)_4$ meson. When $n=4$ is substituted in Eq. (\ref{itp1}) it yields the DCE, underlying the $(D^0)_4$,  equals 1039.93 nat. In this way, this value of the DCE can be replaced on the left-hand side of Eq. (\ref{itq11}),  which can be solved as a function of the $m^2$ variable, immediately yielding the mass  $m_{(D^0)_4}= 3689.27$ MeV. Analogously, Eq. (\ref{itp1}) can be  evaluated for $n=5$, and  the DCE is 1145.50 nat. Hence, Eq. (\ref{itq11}) can be worked out for it, from which $m_{(D^0)_5}= 4126.97$ MeV.  Charmed meson resonances with $n>5$ are configurationally more unstable and, therefore, unlikely to be produced in particle collisions in the current energy range of the existing experiments. These outcomes are listed in Table 
\ref{scalarmasses102}. 
	\begin{table}[H]
\begin{center}\begin{tabular}{||c|c|c|c|c||}
\hline\hline
$n$ & State & $M_{\scalebox{.64}{\textsc{Exp.}}}$ (MeV)  & $M_{\scalebox{.64}{\textsc{AdS/QCD and hybrid}}}$ (MeV)& DCE (nat)  \\
       \hline\Xhline{2\arrayrulewidth}
1 &\;$D^0\;$ &$1864.85\pm0.05$ & $1671.14$  & 728.65 \\ \hline
2$^*$ &\;$D^0(2550)^0\;$ & $2549\pm19$ & $2778$ & 817.89 \\ \hline
3$^*$ &\;$D(3000)^0\;$ & $\;3214\pm29\pm49 \;$ & $3317$ & 939.90 \\ \hline
4& \;$(D^0)_4\;$& ---------  & $3689^\diamond$& 1039.93 \\\hline
5& \;$(D^0)_5\;$& ---------     & $\clt{4126^\diamond}$ & 1145.50 \\\hline
\hline
\end{tabular}
\caption{Table \ref{scalarmasses1} added with the resonances  of the $D^0$ meson family for $n=4,5$. In the 4${}^{\rm th}$ column, the states a `` ${}^\diamond$ '' stand for the mass spectrum extrapolated by employing both Eqs.  (\ref{itp1}, \ref{itq11}), which interpolate the experimental mass spectrum of $D^0$ meson states for $n=1, 2, 3$. } \label{scalarmasses102}
\end{center}
\end{table}
The $(D^0)_5$ strange axial-vector kaon resonance, whose mass has been estimated to equal $(4127\pm67)$ MeV, may match the  $X(4160)$ meson state,  whose experimental mass is $4146\pm18\pm33$ MeV and it is omitted from the summary table listing of mesons in PDG \cite{pdg}.

 \subsection{DCE of charmed $D^*$ meson resonances}
  \label{ndce1}

\begin{figure}[H]
	\centering
	\includegraphics[width=9.8cm]{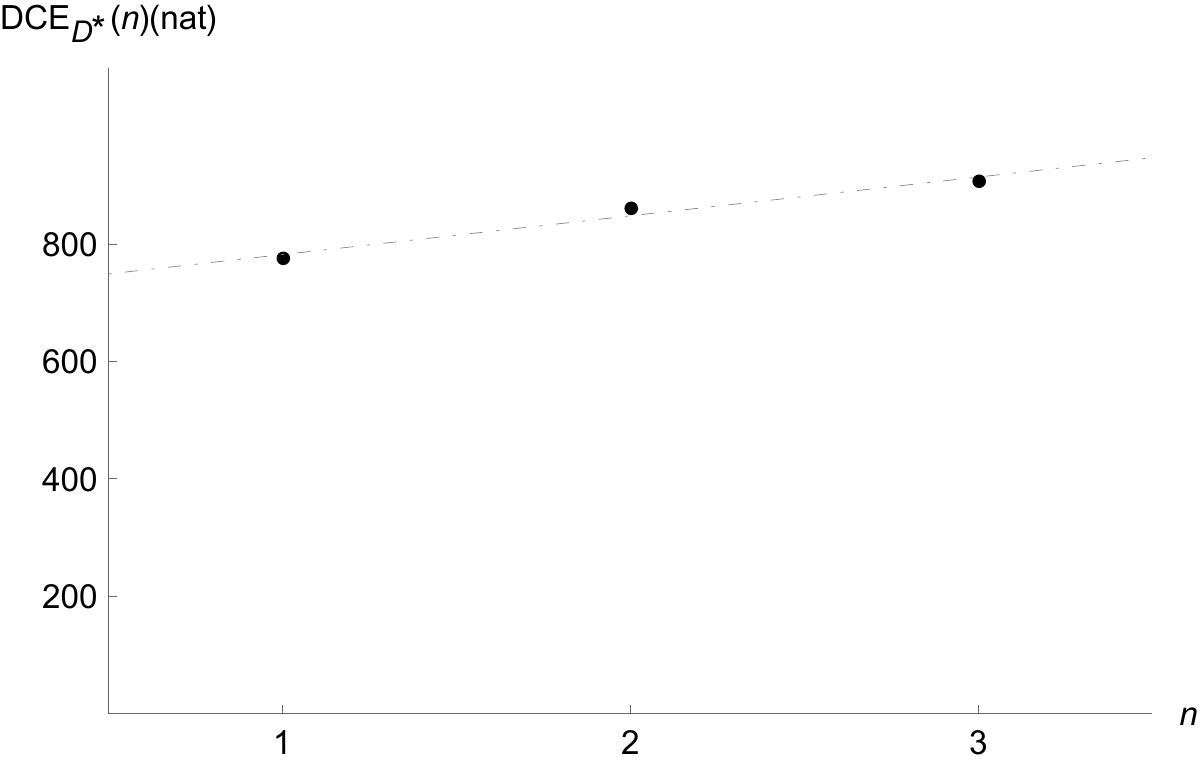}
	\caption{DCE of $D^*$ meson resonances as a function of the principal quantum number $n$, for  $n=1,2,3$ (respectively corresponding to the states $D^*(2007)^0$,  $D^*_1(2600)^0$, $D^*_1(2760)^0$ \cite{pdg}).  
The DCE-Regge trajectory of the first kind, Eq. (\ref{itp1}), is plotted by a dot-dashed line.}
	\label{dcends}
\end{figure}

\clt{Now, the DCE of the $D^*$ meson family will be computed and discussed. The first meson resonance is labeled by  $n=1$, regarding  the 
$D^*(2007)^0$ meson, whereas $n=2$ designates the $D^*_1(2600)^0$ meson state, and $n=3$ corresponds to the $D^*_1(2760)^0$ meson state.  
The DCE  can then be estimated from Eq. \eqref{rhom} for the action functional (\ref{act1}), considering the integrand of the actions (\ref{act2}) - (\ref{act5}). 
Subsequently, the protocol of using Eqs. (\ref{fou1}, \ref{dce1}) has  numerical values  listed in Table \ref{scalarmasses60}.}
\begin{table}[H]
\begin{center}
\begin{tabular}{||c|c|c|c||}
\hline\hline
$n$ & \;{} $D^*$ meson resonances\; & $M_{\scalebox{.64}{\textsc{exp}}}$ (MeV)&DCE (nat) \\
       \hline\Xhline{2\arrayrulewidth}
1 &\;$D^*(2007)^0$\; &$2006.85\pm0.05$ &776.11 \\ \hline
2$^*$ &\;$D^*(2007)^0,D^*_1(2600)^0\;$ & $2627\pm10 $ &859.35 \\ \hline
3 &\;$D^*_1(2760)^0\;$ & $\;2781\pm18\pm13 \;$ & 1002.28 \\ 
\hline\hline
\end{tabular}
\caption{DCE of the $D^*$ meson family, for $n=1, 2, 3$.} \label{scalarmasses60}
\end{center}
\end{table}
The experimental mass spectrum in PDG \cite{pdg} corroborates the predictions of hadronic production of the $D^*$ meson family, reported in Ref.  \cite{Zhao:2016mxc}. 
The DCE-Regge trajectory of the first  kind corresponds to expressing the DCE with respect to the $n$ quantum number. Linear interpolation of data in Table \ref{scalarmasses60} yields the  DCE-Regge trajectory of the first  kind, 
\begin{eqnarray}\label{itp1s}
\!\!\!\!\!\!\!\!\!\!\!\!{{\rm DCE}}_{D^*}(n)&=& 71.085\,n + 709.077,  \end{eqnarray}
within $0.12\%$ RMSE. 
Data in Table \ref{scalarmasses50} and Eq. (\ref{itp1s}) are together depicted in Fig. \ref{dcends}.

\begin{figure}[H]
	\centering
	\includegraphics[width=9.8cm]{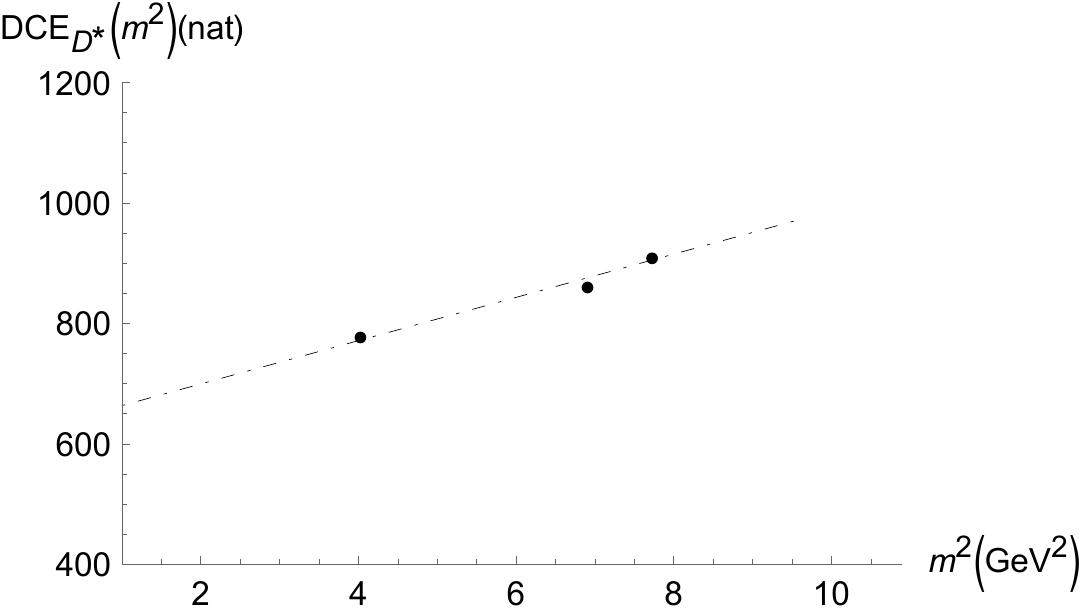}
	\caption{DCE of the $D^*$ meson family displayed as a function of their respective squared mass, for  $n=1,2,3$ (corresponding, respectively, to the $D^*(2007)^0$,  $D^*_1(2600)^0$, $D^*_1(2760)^0$ heavy-light-flavor meson states \cite{pdg}). 
The DCE-Regge trajectory of the second  kind in Eq. (\ref{itq11}) corresponds to the interpolating dot-dashed line.}
	\label{dcedsm2s}
\end{figure}

In addition, the DCE underlying the $D^*$ charmed meson family can be displayed as a function of the experimental (squared) mass spectrum of the heavy-light-flavor $D^*$ charmed meson states, which can be read off Table \ref{scalarmasses2}. Fig. \ref{dcedsm2s} illustrates it, with an interpolation curve corresponding to a DCE-Regge trajectory of the second  kind, given by Eq. (\ref{itq11s}). 
\noindent The DCE-Regge trajectory of the second  kind, expressing the DCE of $D^*$ meson family as a function of their respective (squared) mass spectrum, $m^2$ (GeV${}^2$), is given by the interpolation function 
\begin{eqnarray}
\label{itq11s}
\!\!\!\!\!\!\!\!\!\!\!{\rm DCE}_{D^*}(m^2) \!\!&\!=\!& 35.929 \,m^2 + 627.734,
   \end{eqnarray}   
   within $0.1\%$ RMSE. One could try to reduce the RMSE with a higher-order polynomial interpolation, as a quadratic one, as a first attempt. However, the first reason is the lack of necessity, as the RMSE of the interpolating line \eqref{itq11s} lies within $0.1\%$. It is an unnecessary refinement since the experimental errors involved are, in general, beyond this lower margin. Another reason for keeping the linear interpolation in  Eq. (\eqref{itq11s}) is the existence of three points  in Fig. \eqref{dcedsm2s}. Interpolating the $D^*(2007)^0$,  $D^*_1(2600)^0$, $D^*_1(2760)^0$ states masses with three parameters, required by a quadratic ($\propto m^4$) interpolation, would be overfitted, lacking predictivity, and it might generalize not so appropriately to novel data sets for the $D^*$ light-heavy-flavor meson family.

Eqs. (\ref{itp1}, \ref{itq11}) represent the two types of DCE-Regge trajectories, containing the aspects that are necessary to get the mass spectrum of heavy-light-flavor  resonances in the $D^*$ charmed meson family with $n>3$.  This protocol is established just on the DCE, which represents the configurational entropy underlying the mesons in the 4-flavor AdS/QCD, and on the experimental mass spectrum of the first three states of the $D^*$ meson family in PDG \cite{pdg}. Therefore, it is a more realistic method when compared to the 4-flavor AdS/QCD, where the mass spectrum of $D^*$ resonances can be read off as the eigenvalues in the Schr\"odinger-like equation. 
Nevertheless one must emphasize that Eq.  (\ref{itq11s}), representing the DCE-Regge trajectory of the second kind expressing the DCE of the $D^*$ meson family as a function of their respective (squared) mass spectrum, is acquired by interpolating the experimental mass spectrum of the $D^*$ meson states, also shown in Fig. \ref{dcedsm2s}. In this way, the extrapolation method relies on the experimental data set in PDG \cite{pdg}, rather than using just the AdS/QCD apparatus to predict the mass spectrum of the next generation of  heavy-light-flavor  resonances in the $D^*$ charmed meson family. 
Substituting $n=4$ into Eq. (\ref{itp1s}) yields the DCE corresponding to  the $(D^*)_4$ states equals 980.083 nat. Replacing it on the left-hand side in Eq. (\ref{itq11s}) and solving for the $m^2$ variable, it implies that   $m_{(D^*)_4}= 3180.62$ MeV corresponds the mass of  the $(D^*)_4$ charmed meson state. Eq. (\ref{itp1s}) can then be evaluated for $n=5$, with the DCE equal to 1046.17 nat. Hence Eq. (\ref{itq11s}) can be worked out for it, giving $m_{(D^*)_5}= 3473.34$ MeV as the mass for the $(D^*)_5$ charmed meson resonance.   These results are summarized in Table 
\ref{scalarmasses102}. 
	\begin{table}[H]
\begin{center}\begin{tabular}{||c|c|c|c|c||}
\hline\hline
$n$ & State & $M_{\scalebox{.64}{\textsc{Exp.}}}$ (MeV)  & $M_{\scalebox{.64}{\textsc{AdS/QCD and hybrid}}}$ (MeV)& DCE (nat)  \\
       \hline\Xhline{2\arrayrulewidth}
1 &\;$D^*(2007)^0$\; &$2006.85\pm0.05$ & $2296.33$  & 776.11 \\ \hline
2$^*$ &\;$D^*_1(2600)^0\;$ & $2627\pm10 $ & $2512$ & 859.35 \\ \hline
3 &\;$D^*_1(2760)^0\;$ & $2781\pm18\pm13 $ & $2756$ & 918.28 \\ 
\hline4& \;$(D^*)_4$& ---------  & 3190${}^\diamond$& 993.42 \\\hline
5& \;$(D^*)_5\;$& ---------     & 3486$^\diamond$ & 1064.50 \\\hline
\hline
\end{tabular}
\caption{Table \ref{scalarmasses2} added with the resonances  of the $D^*$ charmed meson family for $n=4,5$. In the 4${}^{\rm th}$ column, the states a `` ${}^\diamond$ '' stand for the mass spectrum extrapolated by employing both Eqs.  (\ref{itp1s}, \ref{itq11s}), which interpolate the experimental mass spectrum of $D^*$ meson states for $n=1, 2, 3$. } \label{scalarmasses102st}
\end{center}
\end{table}
The $(D^*)_4$ meson resonance, whose mass has been estimated to equal ($3190\pm 38$) MeV, might match the  $X(3250)$ meson state, which is a particle in the further meson states section of PDG \cite{pdg} and has experimental mass  $3245\pm8\pm20$ MeV, obtained from 4-body decays.
 Moreover, the $(D^*)_5$ meson resonance has a mass here predicted to be equal ($3486\pm 106$)  MeV. Hence, it may also correspond to the  $X(3350)$ meson resonance reported in the PDG, with experimental mass value $3350^{+10}_{-20}\pm20$ MeV \cite{pdg}. \\

 \subsection{DCE of $\eta_c$ charmonium-like meson resonances}
 \label{ndce2}
 
 \clt{One can evaluate the DCE of the $\eta_c$ charmonium-like meson resonances already detected and reported in the PDG \cite{pdg}, for  $n=1,2$,  respectively for the states 
$\eta_c(1S)$ and $\eta_c(2S)$ \cite{pdg}. 
Superseding Eq. \eqref{rhom} in Eqs. (\ref{fou1}, \ref{dce1}), the DCE of the $\eta_c$ charmonium-like meson family is calculated using numerical methods, displayed in Table \ref{scalarmasses501}}.
\begin{table}[H]
\begin{center}
\begin{tabular}{||c|c|c||}
\hline\hline
$n$ & \;{} $\eta_c$ meson family\; & DCE (nat) \\
       \hline\Xhline{2\arrayrulewidth}
1 &\;$\eta_c(1S)\;$ &  935.11 \\ \hline
2 &\;$\eta_c(2S)\;$ & 1101.36 \\ \hline
\hline
\end{tabular}
\caption{DCE of the states in the $\eta_c$ charmonium-like meson family with $n=1, 2$.} \label{scalarmasses501}
\end{center}
\end{table}
The  DCE-Regge trajectory of the first kind for the $\eta_c$ charmonium-like meson resonances refers to the DCE dependent on the principal quantum number $n$. 
Linear interpolation of data in Table \ref{scalarmasses501} constitutes the DCE-Regge  trajectory of the first kind, given by  
\begin{eqnarray}\label{itp11etac}
\!\!\!\!\!\!\!\!\!\!\!\!\!\!\!\!{{\rm DCE}}_{\eta_c}(n)\!&\!=\!&\!166.25\,n+ 768.86.  \end{eqnarray}

\begin{figure}[H]
	\centering
	\includegraphics[width=10.5cm]{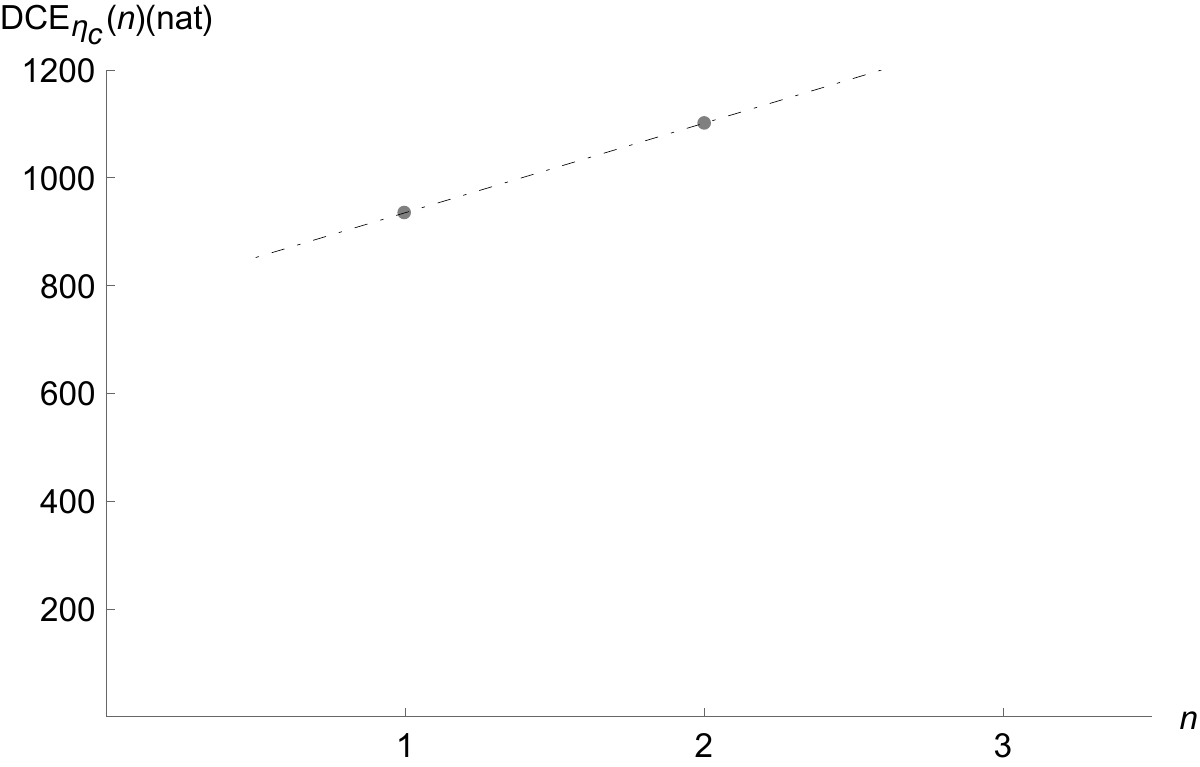}
	\caption{DCE of the $\eta_c$ charmonium-like meson family, for  $n=1,2$ (respectively corresponding to the $\eta_c(1S)$ and $\eta_c(2S)$ meson states  in PDG \cite{pdg}).  
The DCE-Regge trajectory of the first kind (\ref{itp11etac}) is displayed as the interpolating dot-dashed line.}
	\label{cen1dd}
\end{figure}
\begin{figure}[H]
	\centering
	\includegraphics[width=10.5cm]{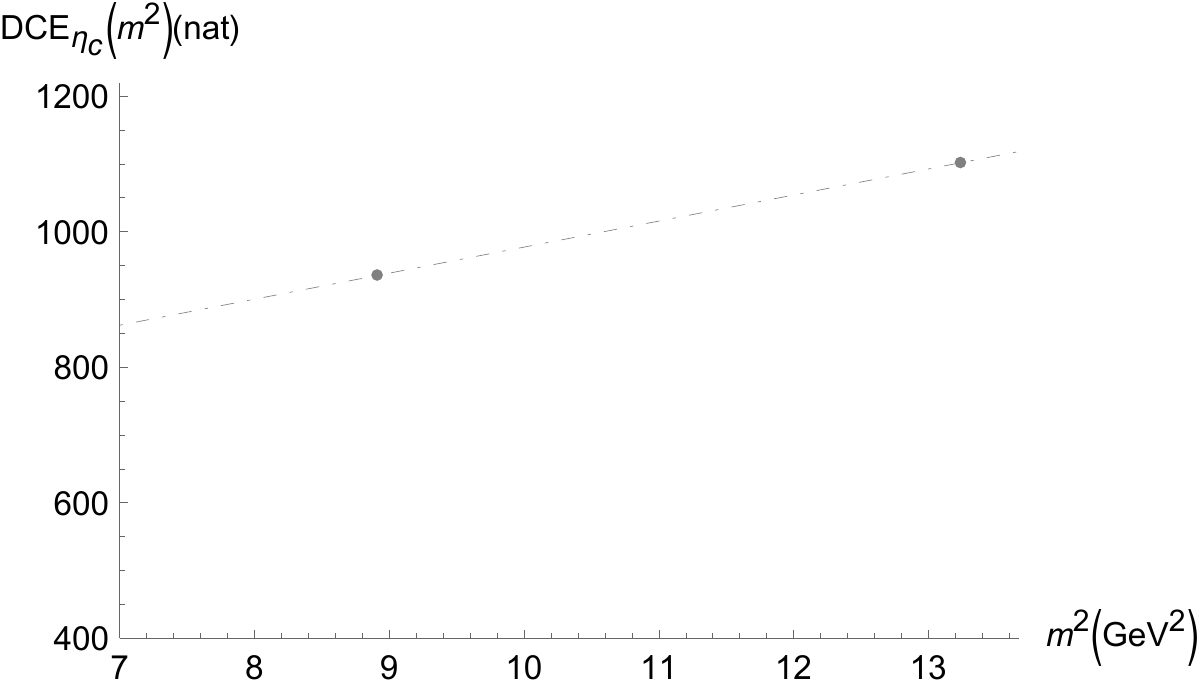}
	\caption{DCE of the $\eta_c$ charmonium-like meson states as a function of their squared mass, for  $n=1,2$ (respectively corresponding to the 
$\eta_c(1S)$ and $\eta_c(2S)$ meson states in PDG \cite{pdg}). 
The DCE-Regge trajectory of the second kind, in Eq. (\ref{itq112etac}), corresponds to the interpolating dot-dashed line.}
	\label{cem12}
\end{figure}

The DCE underlying the $\eta_c$ charmonium-like meson states can also be seen as a function of their experimental mass spectrum, accounting for the DCE-Regge trajectory of the second kind. Taking into account the DCE of the set of states in the $\eta_c$ charmonium-like meson family in Table \ref{scalarmasses501}, it can also be realized as a mass-dependent function for each $\eta_c$ charmonium-like meson resonance, immediately available through Table \ref{scalarmasses3}. 
Fig. \ref{cem12} summarizes these important results. 
Interpolating the  data obtained yields the DCE-Regge trajectory of the second  kind, whose  interpolation expression is represented in Eq. (\ref{itq112etac}). 
\noindent 
The DCE-Regge trajectory of the second kind, which makes the DCE of the $\eta_c$ charmonium-like meson family to be dependent on the squared mass spectrum of these states, $m^2$ (GeV${}^2$), is given by 
\begin{eqnarray}
\label{itq112etac}
\!\!\!\!\!\!\!\!\!\!\!\!{\rm DCE}_{\eta_c}(m^2) \!&\!=\!& 38.4126\,m^2 + 593.052.   \end{eqnarray}     
By the use of Eq. (\ref{itp11etac}), the DCE of the $\eta_c$ charmonium-like meson resonances  can be inferred and these values, for each $n>2$, can be replaced into the left-hand side of the DCE-Regge trajectory of the second kind (\ref{itq112etac}), which can be algebraically solved, yielding the mass spectrum of the next generation of $\eta_c$ charmonium-like meson resonances, relying solely on the experimental mass spectrum of the already observed $\eta_c$ charmonium-like meson resonances reported in PDG \cite{pdg}. The mass of the  $\eta_c(3S)$ resonance can be estimated when $n=3$ is put in Eq. (\ref{itp11etac}), yielding the DCE equal to 1267.61  nat. Hence, this value is substituted in Eq. (\ref{itq112etac}), implying that  $m_{\eta_c(3S)}= 4190.57$ MeV is the estimated mass for the $\eta_c(3S)$ states. To obtain the mass of the $\eta_c(4S)$ resonance, Eq. (\ref{itp11etac}) has to be evaluated with $n=4$, with DCE equal to  1433.86  nat. Therefore Eq. (\ref{itq112etac}) can be worked out, yielding $m_{\eta_c(4S)}= 4668.56$ MeV. These results are condensed  in Table 
\ref{scalarmasses1022}. 
	\begin{table}[H]
\begin{center}\begin{tabular}{||c|c|c|c|c||}
\hline\hline
$n$ & State & $M_{\scalebox{.64}{\textsc{Exp.}}}$ (MeV)  & $M_{\scalebox{.64}{\textsc{AdS/QCD-hybrid}}}$ (MeV)&DCE (nat) \\
       \hline\Xhline{2\arrayrulewidth}
1 &\;$\eta_c(1S)\;$ & $2984.1\pm0.4$ & $2600.8$  & 935.11 \\ \hline
2 &\;$\eta_c(2S)\;$ & $3637.7\pm0.9 $ & $3641.2$&1101.36  \\\hline
3& \;$\eta_c(3S)\;$& ---------     & $4190{}^\diamond$&1267.61   \\\hline
4& \;$\eta_c(4S)\;$&  ---------   & $4668{}^\diamond$&1433.86 \\\hline
\hline
\end{tabular}
\caption{Table \ref{scalarmasses3} completed with $\eta_c$ charmonium-like resonances with $n>3$.  The  masses extrapolated from the DCE-Regge trajectories (\ref{itp11etac}, \ref{itq112etac}), for $n=3,4$, in the 4${}^{\rm th}$ column, are indicated with a `` ${}^\diamond$ ''. } \label{scalarmasses1022}
\end{center}
\end{table}
Extrapolating the DCE-Regge trajectories (\ref{itp11etac}, \ref{itq112etac}) for $n=3$ showed that the mass value of the $\eta_c(3S)$ charmonium-like meson resonance equals 4126.97 MeV,  which might match the  $X(4160)$ meson state with experimental mass $4146\pm18\pm33$ MeV reported in PDG \cite{pdg}. Ref.  \cite{Wang:2016mqb} corroborates these results, which alternatively propose the $X(4160)$ meson resonance as a heavier resonance in the $\eta_c$ family as a new charmonium-like state obtained from the process $e^+e^-\to J/\psi D^* \bar{D}^*$. 
The 2-body open charm  Okubo--Zweig--Iizuka-allowed strong decay of $\eta_c$ resonances can be investigated by an improved
Bethe--Salpeter method combined with the ${}^3
P_0$ model. Lastly, the $\eta_c(4S)$ charmed meson resonance, with $n=4$, has mass 4668.56  MeV obtained using DCE-techniques in the 4-flavor AdS/QCD and might match the  $X(4630)$ meson state with mass  $4626\pm16^{+28}_{-110}$ MeV, experimentally measured and reported in PDG \cite{pdg}.

 \subsection{DCE of $\chi_{c1}$ charmonium-like meson resonances}
 \label{ndce3}
 
\begin{figure}[H]
	\centering
	\includegraphics[width=9.8cm]{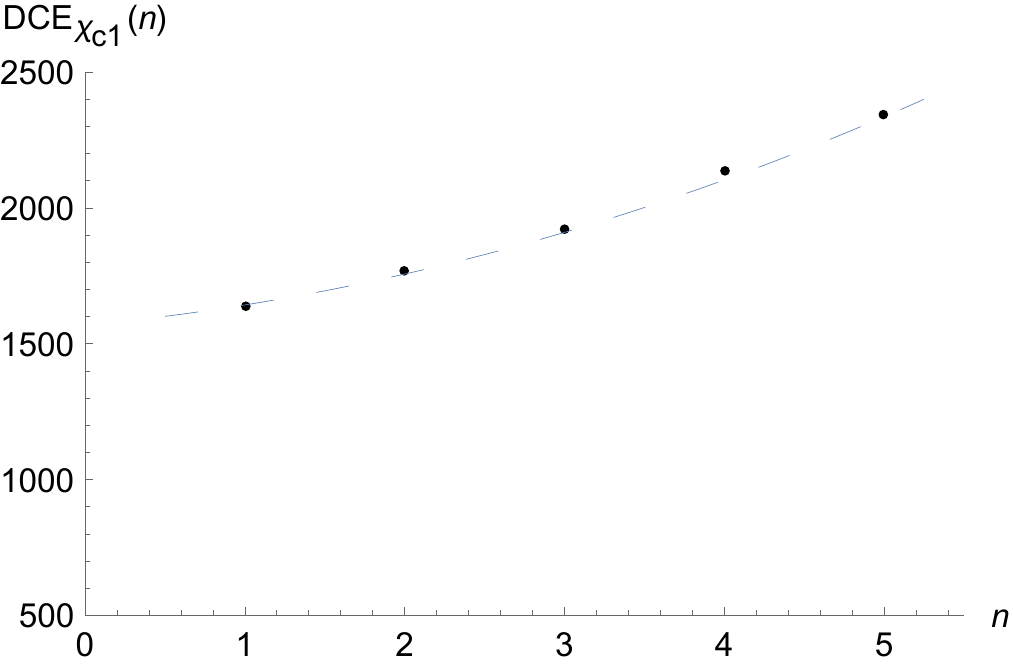}
	\caption{DCE of the $\chi_{c1}$ charmonium-like meson family, for  $n=1,\ldots, 5$ (respectively corresponding to the 
$\chi_{c1}(1P)$, $\chi_{c1}(3872)$, $\chi_{c1}(4140)$, $\chi_{c1}(4274)$, and $\chi_{c1}(4685)$ in PDG \cite{pdg}).  
The DCE-Regge trajectory of the first kind (\ref{itp11c1}) is displayed as the interpolating long-dashed line.}
	\label{cen1ddc}
\end{figure}

The $\chi_{c1}$ meson family can be now discussed. 
One can evaluate the DCE of the $\chi_{c1}$ charmonium-like  meson resonances already detected and reported in the PDG \cite{pdg}, for  $n=1,\ldots,5$,  respectively representing the states $\chi_{c1}(1P)$, $\chi_{c1}(3872)$, $\chi_{c1}(4140)$, $\chi_{c1}(4274)$, $\chi_{c1}(4685)$. 
Using the energy density \eqref{rhom} into Eqs. (\ref{fou}, \ref{dce_modal}, \ref{dce}) or, equivalently, Eqs. (\ref{fou1}, \ref{dce1}), the DCE underlying  the $\chi_{c1}$ meson family is numerically computed and listed in Table \ref{scalarmasses502}.
\begin{table}[H]
\begin{center}
\begin{tabular}{||c|c|c||}
\hline\hline
$n$ & \;{} $\chi_{c1}$ meson family\; & DCE (nat) \\
       \hline\Xhline{2\arrayrulewidth}
1 &\;$\chi_{c1}(1P)\;$ & $1675.51$  \\ \hline
2 &\;$\chi_{c1}(3872)\;$ & $1798.19$ \\ \hline
3 &\;$\chi_{c1}(4140)\;$ & $1958.78$  \\ \hline
4 &\;$\chi_{c1}(4274)\;$ & $2139.44$ \\ \hline
5* &\;$\chi_{c1}(4685)\;$ & $2346.01$ \\ \hline 
\hline
\end{tabular}
\caption{DCE of the states in the $\chi_{c1}$ charmonium-like meson family with $n=1, \ldots, 5$.} \label{scalarmasses502}
\end{center}
\end{table}
\noindent The  DCE-Regge trajectory of the first kind for the $\chi_{c1}$ charmonium-like mesons,  displays the DCE as a function dependent on $n$, given by 
\begin{eqnarray}\label{itp11c1}
\!\!\!\!\!\!\!\!\!\!\!\!\!\!\!\!{{\rm DCE}}_{\chi_{c1}}(n)\!&\!=\!&\! 19.8464\, n^2+ 54.1464 \,n + 1568.84,  \end{eqnarray}
 within $0.01\%$ RMSE. 
Fig. \ref{cen1ddc} arrays corresponding results, whose quadratic polynomial  interpolation takes into account data in Table \ref{scalarmasses502}.

\begin{figure}[H]
	\centering
	\includegraphics[width=11.5cm]{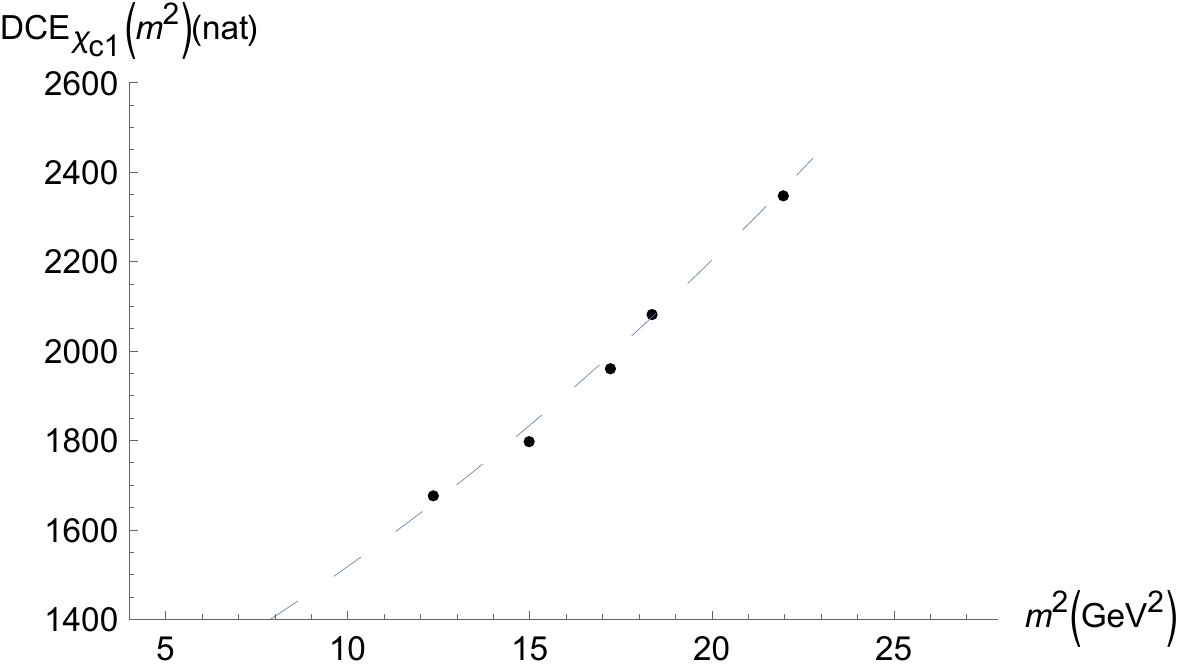}
	\caption{DCE of the $\chi_{c1}$ meson states as a function of their squared mass, for  $n=1,\ldots,5$ (respectively corresponding to the 
$\chi_{c1}(1P)$, $\chi_{c1}(3872)$, $\chi_{c1}(4140)$, $\chi_{c1}(4274)$, and $\chi_{c1}(4685)$ meson states,  in PDG \cite{pdg}). 
The DCE-Regge trajectory of the second  kind, in Eq. (\ref{itq112c1}), corresponds to the interpolating long-dashed line.}
	\label{cem12c1}
\end{figure}

The DCE underlying the $\chi_{c1}$ meson states can be understood as a function of the experimental mass spectrum of these meson states, accounting for the DCE-Regge trajectory of the second kind, as illustrated in Fig. \ref{cem12c1}. Interpolation yields the DCE-Regge trajectory of the second kind, whose formula is presented in Eq. (\ref{itq112c1}). 
\noindent The DCE-Regge trajectory of the second  kind, which makes the DCE of the $\chi_{c1}$ meson family dependent on the squared mass spectrum of these states, $m^2$ (GeV${}^2$), is given by 
\begin{eqnarray}
\label{itq112c1}
\!\!\!\!\!\!\!\!\!\!\!\!{\rm DCE}_{\chi_{c1}}(m^2) \!&\!=\!&1.06344\, m^4+ 36.5495\, m^2 + 1046.29, 
   \end{eqnarray} within $2.1\%$  RMSE.  Quadratic interpolation\footnote{Here by ``quadratic'' we mean quadratic with respect to the parameter $m^2$, namely, up to $m^4$.},  
 is neither overfitted nor underfitted, being well suited for numerical inference with optimal predictivity, using three parameters to describe the five  
  $\chi_{c1}(1P)$, $\chi_{c1}(3872)$, $\chi_{c1}(4140)$, $\chi_{c1}(4274)$, and $\chi_{c1}(4685)$ meson resonances.    
To calculate the value of the mass of the  $(\chi_{c1})_6$ resonance, one first selects $n=6$ in Eq. (\ref{itp11c1}), which yields the DCE equals 2608.19  nat. This value is hence replaced in Eq. (\ref{itq112c1}) and produces the mass $m_{(\chi_{c1})_6}= 4981.27$ MeV. For finding the mass of the $(\chi_{c1})_7$ resonance, Eq. (\ref{itp11c1}) must be evaluated with $n=7$, originating the DCE value 2920.34 nat. Therefore Eq. (\ref{itq112c1}) can be solved for the mass value $m_{(\chi_{c1})_7}= 5307.81$ MeV. Finally, Eq. (\ref{itp11c1}) takes into account the $n=8$ resonance can be considered in Eq. (\ref{itp11c1}), giving the DCE of the $(\chi_{c1})_8$ to equal 3272.18 nat. When Eq. (\ref{itq112c1}) is solved for it yields $m_{(\chi_{c1})_8}= 5628.83$ MeV. These results are summarized and displayed in Table 
\ref{scalarmasses1022c1}. 
	\begin{table}[H]
\begin{center}\begin{tabular}{||c|c|c|c|c||}
\hline\hline
$n$ & State & $M_{\scalebox{.64}{\textsc{Exp.}}}$ (MeV)  & $M_{\scalebox{.64}{\textsc{AdS/QCD and hybrid}}}$ (MeV)& DCE (nat) \\
       \hline\Xhline{2\arrayrulewidth}1 &\;$\chi_{c1}(1P)\;$ & $\;3510.67\pm0.05\;$ & $\;3464.72\;$  & 1675.51
 \\ \hline
2 &\;$\chi_{c1}(3872)\;$ & $3871.64\pm0.06$ & $3808.17$  &  1798.19 \\ \hline
3 &\;$\chi_{c1}(4140)\;$ & $4146.5\pm3.0$ & $4138.1$  & 1958.78 \\ \hline
4 &\;$\chi_{c1}(4274)\;$ & $4286^{+8}_{-9}$ & $4460$  &  2139.44
 \\ \hline
5* &\;$\chi_{c1}(4685)\;$ & $4684\pm7^{+13}_{-16}$ & $4723$  & 2346.01\\ \hline 
6& \;$(\chi_{c1})_6\;$& ---------     & $4981{}^\diamond$   &2608.19\\\hline
7& \;$(\chi_{c1})_7\;$&  ---------   & $5307{}^\diamond$&2920.34 \\\hline
8& \;$(\chi_{c1})_8\;$&  ---------   & $5628{}^\diamond$&3272.18 \\\hline
\hline
\end{tabular}
\caption{Table \ref{scalarmasses4} completed with $\chi_{c1}$ resonances with $n>5$.  The  masses extrapolated from the DCE-Regge trajectories (\ref{itp11c1}, \ref{itq112c1}), for $n=6,7,8$, in the 4${}^{\rm th}$ column, are indicated with a `` ${}^\diamond$ ''. The errors were propagated to the model predictions and displayed for the new $\chi_{c1}$ resonances, for $n=6, 7,8$} \label{scalarmasses1022c1}
\end{center}
\end{table}
\noindent Table \ref{scalarmasses1022c1} provides an important database for 
searching heavier resonances in the $\chi_{c1}$ charmonium-like family. 
The lightest resonances in this family are produced in high-energy proton-proton collisions at large transverse momenta, with production  rates comparable to the quarkonium
$\psi(2S)$. The $\chi_{c1}(3872)$ charmonium-like state was shown to  be generated from string-breaking phenomena   \cite{Ikeno:2019grj,Bruschini:2021fuw}.

\section{Conclusions}\label{iv}

The DCE underlying the 4-flavor AdS/QCD was hybridized with 
phenomenological data regarding the experimental mass spectra  of 
heavy-light-flavor meson families and charmonium-like families, respectively composed by the charmed $D^0$, $D^0(2550)$, and $D^0(3000)$ mesons, the  $D^*(2007)^0$, $D^*_1(2600)^0$, and $D^*_1(2760)^0$ mesons,  the $\eta_c(1S)$ and $\eta_c(2S)$ meson states, and the $\chi_{c1}(1P)$,  $\chi_{c1}(3872)$, $\chi_{c1}(4140)$, $\chi_{c1}(4274)$, and $\chi_{c1}(4685)$, reported in the PDG \cite{pdg}. 
This technique permitted to predict the mass spectrum of heavier charmed resonances with higher principal quantum numbers in all these four charmed meson families. 
\clt{This method involving the DCE takes into account 
directly the mass spectrum of each meson family here reported, to predict 
the mass spectrum of heavier meson resonances in each meson family addressed, besides already taking into account the experimental masses through the fit of the model parameters. Our approach is complementary, also involving the information entropy content in meson families. 
Therefore, there are two possible databases for the mass spectra  of 
heavy-light-flavor meson families and charmonium-like families, in the AdS/QCD setup. The first one is the meson mass spectra as the squared eigenvalue of the Schrödinger-like equations, whereas the second one consists of the mass spectra extrapolated from the DCE-
Regge trajectories, where we compute the DCE underlying the 4-flavor AdS/QCD and hybridize it with phenomenological data regarding the experimental mass spectra of heavy-light-flavor meson families and charmonium-like families reported in the PDG \cite{pdg}. Instead of obtaining the meson mass spectra as the squared eigenvalue of the Schrödinger-like equations, we take into account the energy density underlying the system, namely, the temporal component of the energy-momentum tensor. This approach is also phenomenologically robust, relying on the experimental meson mass spectra in PDG \cite{pdg}}.  All the possibilities for matching the next generation of heavier charmed meson resonances to further meson states in particle listings of the PDG were analyzed and discussed. 
 In addition, DCE-based protocols are more concise and present the advantage of considering phenomenological physical aspects. In fact, 
 the next generation of charmed meson states is obtained through the extrapolation of the DCE-Regge trajectories, whose second kind takes into account the mass spectrum of the heavy-light-flavor $D^0$ and $D^*$ meson families, as well as the $\eta_c$ and $\chi_{c1}$  charmonium-like meson families, experimentally measured and reported in the PDG. The hybrid method involving the use of the DCE establishes a method to estimate the mass spectra of heavier charmed meson resonances in the $D^0$, $D^*$, $\eta_c$, and $\chi_{c1}$ charmed meson families, based on the extrapolation of experimental data in PDG \cite{pdg}. This procedure provides a theoretical roadmap to identify newly detected meson states, also matching theoretical predictions to the already detected particles that remain detached from any specific meson family. 

Table \ref{scalarmasses102} gives the mass spectrum of two heavier resonances in the $D^0$ meson family. The $(D^0)_5$ resonance, whose mass has been estimated to equal $(4127\pm59)$ MeV, can match the  $X(4160)$ meson state,  whose experimental mass is $4146\pm18\pm33$ MeV \cite{pdg}. For the  $D^*$ heavy-light-flavor charmed meson family, data summarized in Table  \ref{scalarmasses102st} supplies an important and robust phenomenological database for searching and producing heavier meson resonances in this family.  The $(D^*)_4$ meson resonance, whose mass has been estimated to equal ($3190\pm 38$) MeV, might match the  $X(3250)$ meson state, whose experimental mass is $3245\pm8\pm20$ MeV. Besides, Table \ref{scalarmasses1022} displays the mass spectrum of the $\eta_c$ charmonium-like meson family and it was obtained from extrapolating the DCE-Regge trajectories (\ref{itp11etac}, \ref{itq112etac}) for $n=3$ and $n=4$. It yielded the mass of the $\eta_c(3S)$ meson resonance to be equal to 4126.97 MeV, which may match the  $X(4160)$ meson state,  whose experimental mass is $4146\pm18\pm33$ MeV. These results  corroborate with Ref. \cite{Wang:2016mqb}, which also proposed the $X(4160)$ meson resonance as a heavier resonance in the $\eta_c$ family as a new charmonium-like state. Also, the $\eta_c(4S)$ charmed meson resonance, with $n=4$, has mass 4668.56  MeV obtained using DCE-techniques in the 4-flavor AdS/QCD and may  match the  $X(4630)$ meson state in PDG \cite{pdg}, with mass  $4626\pm16^{+28}_{-110}$ MeV experimentally measured and reported in PDG \cite{pdg}. Finally, Table \ref{scalarmasses1022c1} provides a useful database for investigating heavier meson resonances in the $\chi_{c1}$ charmonium-like meson family. Some candidates for the next generation of charmed meson resonances, already reported in PDG as further meson states and omitted from the summary table, have been  matched to our results, and some other heavier resonances provide a reliable database that might guide efforts for sifting heavier charmed meson  resonances which have not been observed hitherto.

\medbreak
\subsubsection*{Acknowledgments} 
 The work of AEB is supported by The S\~ao Paulo Research Foundation (FAPESP) 
(Grant No. 2023/00392-8) and the National Council for Scientific and Technological Development -- CNPq (Grant
No. 301485/2022-4). WdP acknowledges the partial support of CNPq (Grant No. 313030/2021-9) and  CAPES (Grant No. 88881.309870/2018-01). RdR~thanks FAPESP
(Grants No. 2021/01089-1 and No. 2024/05676-7), and CNPq (Grants No. 303742/2023-2 and No. 401567/2023-0), for partial financial support. The authors thank Bruno P. Toniato for fruitful discussions. 
\bibliography{bibliography}

\end{document}